\documentclass[a4paper,11pt]{article}
\pdfoutput=1
\usepackage{jheppub}
\usepackage[T1]{fontenc}
\usepackage{diagbox}
\usepackage{adjustbox}
\usepackage{multirow}
\usepackage{pifont}
\usepackage{array}
\usepackage{mathtools}
\usepackage{booktabs}
\usepackage[dvipsnames]{xcolor}
\usepackage{tcolorbox}
\usepackage{float}
\usepackage{placeins}
\usepackage{braket}

\newcommand{\eqal}[1]{\begin{align}#1\end{align}}

\RequirePackage[normalem]{ulem}

\title{Custodial Symmetry Violation in Scalar Extensions of the Standard Model
}

\author[a]{Huayang Song,}
\author[b]{Xia Wan,}
\author[a,c,d,e,f]{and Jiang-Hao Yu}

\affiliation[a]{CAS Key Laboratory of Theoretical Physics, Institute of Theoretical Physics, Chinese Academy of Sciences, Beijing 100190, P.R.China}
\affiliation[b]{School of physics and Information Technology, Shaanxi Normal University,
Xi'an 710119, China}
\affiliation[c]{School of Physical Sciences, University of Chinese Academy of Sciences,   Beijing 100049, P.R. China}  
\affiliation[d]{Center for High Energy Physics, Peking University, Beijing 100871, China} 
\affiliation[e]{School of Fundamental Physics and Mathematical Sciences, Hangzhou Institute for Advanced Study, UCAS, Hangzhou 310024, China}  
\affiliation[f]{ International Centre for Theoretical Physics Asia-Pacific, Beijing/Hangzhou, China}

\emailAdd{huayangs@itp.ac.cn}
\emailAdd{wanxia@snnu.edu.cn}
\emailAdd{jhyu@itp.ac.cn}

\abstract{The new measurement of the $W$ boson mass from the CDF collaboration shows a significant tension with the Standard Model prediction, which evidences violation of custodial symmetry in the scalar sector. We study the scalar extensions of the Standard Model, which can be categorized into two classes, scalar sector with custodial symmetry (Georgi-Machacek
model and its generalizations) and scalar sector without custodial symmetry, and explore how these extensions fit to the electroweak precision data and the CDF new $m_W$. The favored oblique parameters are coming from either the large mass splitting in the multiplet via the loop contribution or the large vacuum expectation value which breaks custodial symmetry at the tree level. In particular, we find that $\mathcal{O}(100)$ GeV new particles are allowed in the scalar extension scenarios.}

\begin{document} 
\maketitle
\flushbottom

\section{Introduction \label{sec:intro}}
Recently CDF collaboration announced their new measurement of the mass of the $W$ boson with a value of $m_W=80.4335\pm0.0094~{\rm GeV}$~\cite{CDF:2022hxs}. Such value indicates a $7\sigma$ deviation from the Standard Model (SM) prediction $m_W^{SM}=80.357\pm0.006~{\rm GeV}$~\cite{ParticleDataGroup:2020ssz, deBlas:2021wap, CDF:2022hxs}.
Meanwhile, it is also in tension with the previous experimental measurements
including the most precise one from the ATLAS experiment $m_W^{\rm ATLAS}=80.370\pm0.019~{\rm GeV}$~\cite{ATLAS:2017rzl} at $\sim 3\sigma$. Though the discrepancy might be caused by some unknown experimental or theoretical uncertainties, which requires comparative studies by other experiments and also theorists, it could be a hint for new physics and a lot of efforts have been made to provide an explanation under this direction~\cite{Gu:2022htv,Bagnaschi:2022whn,Fan:2022yly,Chen:2022ocr,Du:2022brr,Mondal:2022xdy,Cheng:2022aau, deBlas:2022hdk, Fan:2022dck, Zhu:2022tpr, Lu:2022bgw, Strumia:2022qkt, Liu:2022jdq, Song:2022xts, Babu:2022pdn, Ghorbani:2022vtv, Benbrik:2022dja, Botella:2022rte, Kim:2022xuo, Benincasa:2022elt, Asadi:2022xiy, Bahl:2022xzi, DiLuzio:2022xns, Athron:2022isz, Sakurai:2022hwh, Heo:2022dey, Ahn:2022xax, Paul:2022dds, Endo:2022kiw, Balkin:2022glu, Kanemura:2022ahw, Lee:2022gyf, Tang:2022pxh, Blennow:2022yfm, Cheng:2022jyi, Cacciapaglia:2022xih, Du:2022pbp, Cheung:2022zsb, Crivellin:2022fdf}.

It is well-known that there is a residual global $SU(2)$ symmetry of the Higgs potential in the SM. The Higgs doublet potential is invariant under an $SO(4)\cong SU(2)_L\times SU(2)_R$ symmetry, while the Higgs vacuum expectation value (VEV) only breaks it down to diagonal $SU(2)_V$ subgroup. Such a symmtery guarantees the tree-level value of the Veltman $\rho$ parameter~\cite{Veltman:1977kh} to be unity 
\begin{align}
    \rho\equiv\frac{m_W}{m_Z \cos\theta_W}=1|_{\text{tree}}+\cdots \label{eq:rho}
\end{align} 
where $\cos\theta_W$ is the Weinberg mixing angle. Although the hypercharge $U(1)_Y$ and Yukawa interactions explicitly breaks the custodial symmetry $SU(2)_V$, their effects only come in at loop-level and only result in small deviations from unity at sub-percentage level. Given the previous measurements in electroweak (EW) sector consistent with the SM predictions, beyond the Standard Model (BSM) extensions with new scalars usually assumes that the custodial symmetry $SU(2)_V$ is preserved in the scalar sector. For general $k$ multiplet scalars, the tree-level formula of $\rho$ is shown as~\cite{Langacker:1980js,Gunion:1989we}
\begin{align}
   \rho^{\rm tree} =\frac{\sum_k\eta_k\left[2j_k\left(j_k+1\right)-Y_k^2/2\right]\langle \phi_k \rangle^2}
   {\sum_k Y_k^2\langle \phi_k \rangle^2}
   \label{eq:rhotree}
\end{align}
where $\braket{\phi_k}$ is the VEV of the scalar field $\phi_k$, $j_k$ is the weak-isospin and $Y_k$ is its hypercharge, related to the electric charge by $Q_k=j_{k,3}+Y_k/2$. Here we assume that the real representations only have vanishing hypercharge, and therefore $\eta_k$ is equal to $1$($1/2$) for a complex(real) representation. The tree-level relation $\rho^{\rm tree}=1$ can be automatically fulfilled for scalars with quantum numbers satisfying the condition $(2j_k+1)^2-3Y_k^2=1$ no matter whether the extra scalars develop VEVs or not.

However, the newly released CDF II $W$ mass indicates an obviously custodial symmetry violation (CSV) at a certain level. One can parameterize the effects of new physics phenomena in the EW sector
using oblique parameters $S$, $T$ and $U$~\cite{Peskin:1990zt, Peskin:1991sw, Barbieri:2004qk}, and a latest global fit of EW precision observables (EWPOs) including the $m_W$ anomaly shows that a large $T$ value is favored for $ST$ model or a large $U$ parameter is favored for $STU$ model~\cite{deBlas:2022hdk}. As both $T$ and $U$ parameters violate custodial symmetry, it is time to re-examine the CSV models and the effects of custodial symmetry on the EWPOs and $W$ boson mass. Observing the fact that fermionic and vector bosonic extensions of the SM only generates CSV at loop-level\footnote{A massive $U(1)$ gauge boson who has kinetic mixing with SM $Z$ boson can lead to oblique parameters at tree level. However, we do not consider this case here since such mixing is usually generated from integrating out the heavy particles running in the loop or requires extra scalar particles. Therefore it can either be identified as loop effects or extra augment of scalar extensions.}, we focus on the scalar extensions in this work to compare the CSV effect from the tree and loop levels, e.g. the VEV parameterizing the CSV is zero or not.

The paper is organized as follows. In Sec.~\ref{sec:SExt&CS}, we discuss the most general scalar extensions of the SM. Based on the symmetry property of the scalar potential, they can be categorized into two classes, scalar sector with custodial symmetry and scalar sector without custodial symmetry. In Sec.~\ref{sec:SExt/wCS}, we introduce single scalar extensions of the SM, and perform EW precision fit by directly calculating their $S$, $T$ and $U$ parameters to check if they
could help reduce the tension between the SM EW fit and the new CDF II $m_W$ measurement. The preferred new regions of parameter space are shown and the effects of the extra scalar VEV are also discussed, that without a VEV the custodial symmetry can still be preserved in the Higgs doublet. We analyze the Georgi-Machacek (GM) model in Sec.~\ref{sec:GM} as an example for custodial symmetric scalar extensions. The loop effects could alleviate the tension with few hundreds GeV new scalars. Finally in Sec.~\ref{sec:con}, we conclude.

\section{Scalar sector extensions of the SM and custodial symmetry \label{sec:SExt&CS}}
We are going to consider models that extend the SM scalar sector with the addition of general $SU(2)_L$ multiplets. Unitarity of scattering amplitudes involving pairs of scalars and pairs of $SU(2)_L$ gauge bosons constrains a complex scalar to have weak isospin $j\leq 7/2$ and a real one to have $j\leq 4$, and scatterings involving $U(1)_Y$ gauge boson also put an upper limit on hypercharge $Y$ depending on the size of multiplet~\cite{Hally:2012pu}. These constraints become stronger if additional multiplets are added. Further to require no non-trivial electric charge ($Q=j_3+Y/2$), there are only a finite number of scalar multiplets which can be self-consistent. 

As mentioned in Sec.~\ref{sec:intro}, the relation $\rho^{\rm tree}=1$ can be automatically satisfied under the condition
\begin{align}
    (2j_k+1)^2-3Y_k^2=1
\end{align}
The only possibility beyond $j=1/2, Y=1$ is a representation with weak isospin $j=3$ and $Y=4$. Therefore, the scalar sector with arbitray numbers of doublet with $Y=1/2$ and septet with $Y=4$ automatically preserves the custodial symmetry in their scalar potential~\cite{Hisano:2013sn, Kanemura:2013mc}. The simplest model in this class is the two-Higgs-doublet model (2HDM), which has been discussed to explain the CDF-II $W$ mass anomaly~\cite{CDF:2022hxs} and the newly oblique parameters fitting values~\cite{deBlas:2022hdk} in Ref~\cite{Fan:2022dck, Zhu:2022tpr, Lu:2022bgw, Strumia:2022qkt, Liu:2022jdq, Song:2022xts, Babu:2022pdn, Ghorbani:2022vtv, Benbrik:2022dja, Botella:2022rte, Kim:2022xuo, Benincasa:2022elt, Bahl:2022xzi, Ahn:2022xax, Lee:2022gyf}. We will not discuss this case and refer the readers to these references for further details\footnote{We note that Ref.~\cite{Asadi:2022xiy} makes a statement that a scalar doublet extension of SM can not explain the observed anomaly in $M_W$ measurements. We will not comment on their statement.}. The septet model possesses highly charged scalars ($Q$ upto 5) which could have unique (unusual) collider signatures~\cite{Ramsey-Musolf:2021ldh}. It also predicts a negative $S$, which is favored by the $W$ mass but might contradict the $ST$ parameter fit assuming negligible $U$~\cite{Hisano:2013sn, Kanemura:2013mc}. Therefore, we leave the case with scalar septet for future study. 

In the SM, the custodial symmetry $SU(2)_V$ can be identified as the residual diagonal subgroup of an accidental global $SU(2)_L\times SU(2)_R$ symmetry after EW symmetry breaking. It is, in principle, possible to have a scalar sector with multiple N-plets preserving custodial $SU(2)_V$ if imposing the global $SU(2)_L\times SU(2)_R$ symmetry on the entire scalar sector. These are the Georgi-Machacek (GM) model~\cite{Georgi:1985nv,Chanowitz:1985ug} and its generalizations~\cite{Galison:1983qg, Robinett:1985ec, Logan:1999if, Chang:2012gn, Logan:2015xpa}, which lead to $\rho=1$ at tree-level.

Apart from the above two classes, general scalar extensions of the SM do not possess a symmetry in their scalar sector to preserve $\rho=1$ at tree-level without tuning of parameters. To simplify our discussion, we will only consider single scalar extensions for the scenario where there is no custodial symmetry in the scalar sector.

\section{General single scalar multiplet extensions, custodial symmetry breaking models  \label{sec:SExt/wCS}} 

Though there is no custodial symmetry to preserve $\rho=1$ at tree-level in single scalar extensions of the SM, the SM doublet itself still has a custodial symmetry as long as the additional scalar has a vanishing VEV, and the SM tree-level prediction of $\rho$ parameter is kept as shown
\begin{align}
    \rho^\text{tree}-1\simeq\left\{\eta\left[j\left(j+1\right)-Y^2\right]-2Y^2\right\}\left[4\frac{\braket{\phi}^2}{v^2}+\mathcal{O}\left(\frac{\braket{\phi}^4}{v^4}\right)\right].
\end{align}
A VEV of scalar multiplet can contribute differently to the mass terms of the four components of the SM doublet in Feynman–'t Hooft gauge, which can be understood as an explicit breaking of the custodial symmetry of the SM doublet potential.

Hence to make our discussion more clear, we will consider two cases in this section, \textbf{Case A} where no further VEV is developed except the SM Higgs doublet and only loop contributions can shift $\rho$, and \textbf{Case B} where a VEV is developed for the additional scalar multiplet breaking the custodial symmetry of the SM doublet. We especially use a real triplet and a complex triplet extensions as our examples for the study of \textbf{Case B}. 

\subsection{\textbf{Case A}: General $SU(2)_L$ $N$-plet scalar with $\mathbb{Z}_2$ symmetry\label{sec:Nplet}}
A global $U(1)$ or $\mathbb{Z}_2$ symmetry arise accidentally in the renormalizable level scalar potential, which could protect the scalars from VEVs. Assuming that the parameters of the scalar potential are chosen to respect such symmetry without spontaneously breaking, there is no VEV being developed and the lightest particle is thus forced to be stable, which might be a possible dark matter candidate~\cite{Cirelli:2005uq, Hambye:2009pw, AbdusSalam:2013eya, Pilkington:2016erq, Chao:2018xwz}.

The most general forms of renormalizable scalar potential are discussed in detail in Ref.~\cite{Pilkington:2016erq, Chao:2018xwz, Ramsey-Musolf:2021ldh}, and there are usually 3 free parameters for $\phi$-$H$ Higgs portal interactions depending on the quantum number of the multiplet in the $\mathbb{Z}_2$ symmetric case. However we note that one of these parameters only introduces extra scalar states mixing. For simplification, we do not take into account such interaction, and keeping such term will not significantly modify our results.

The most general gauge-invariant potential satisfying our requirements can be universally expressed as
\begin{align}
    V=m^2 H^\dagger H+\lambda\left(H^\dagger H\right)^2+m_\phi^2\phi^\dagger\phi+\lambda_1 H^\dagger H\phi^\dagger\phi+\lambda_2 H^\dagger\tau^a H\phi^\dagger T^a\phi+\mathcal{O}\left(\phi^4\right)
\end{align}
where $\tau^a$ and $T^a$ are the generators of $SU(2)_L$ in the doublet and $N$-plet representations, respectively. The $\mathbb{Z}_2$ symmetry (or global $U(1)$ symmetry) forbids any terms with odd numbers of one of the fields, and there is no VEV being developed for the $N$-plet field $\phi$. 

The mass of particle $\phi^Q$ with charge $Q=j_3+Y/2$ is given by
\begin{align}
    m_{\phi^Q}^2=m_\phi^2+\frac{v^2}{2}\left(\lambda_1-\frac{1}{2}\lambda_2 j_3\right)\equiv m_{\phi^0}^2-\frac{1}{4}\lambda_2 v^2 Q
\end{align}
where $v\simeq 246$ GeV is the SM Higgs VEV and we define the neutral particle mass
\begin{align}
    m_{\phi^0}^2\equiv m_\phi^2+\frac{v^2}{2}\left(\lambda_1+\frac{1}{4}\lambda_2 Y\right)
\end{align}
Provided that the stable lightest particle $\phi^0$ is neutral, $\lambda_2<0$ is required. Though we also consider the case where the lightest particle can be charged, we can still stick to a negative value of $\lambda_2$ without loss of generality. Further note that in the case of real multiplet (e.g. real triplet), the reality of the multiplet forbids the presence of term proportional to $\lambda_2$~\cite{Hambye:2009pw, Chao:2018xwz, Ramsey-Musolf:2021ldh}, which provides a degenerate spectrum at tree-level, can hardly explain the observed $W$ mass anomaly at this level. However, we see that the loop effects could evade this issue and leave some valid parameter space, which though are not accessible in the running or planned experiments (see upper right panel in Fig.~\ref{plot:realandcomplex}). 

The dominant constraints from electroweak precision observables (EWPOs) arise in the oblique parameters $S$, $T$ and $U$~\cite{Peskin:1990zt, Peskin:1991sw} in a large set of BSM models known as ``universal'' models, in which deviations from the SM reside only in the self-energies of the vector bosons as long as the new states only couple to the SM fermions via $SU(2)_L\times U(1)_Y$ currents~\cite{Barbieri:2004qk}\footnote{In the Standard Model Effective Field Theory (SMEFT) language, ``universal'' theories describes the theories which only involve bosonic operators after matching onto the SMEFT upto field redefinitions. At the $dim$-6 level, ``universal'' theories are completely characterized by 16 parameters including 5 commonly-adopted oblique parameters, $S$, $T$, $W$, $Y$ and $Z$~\cite{Wells:2015uba}. However, in our discussion, we do not use an effective field description but keep the full one-loop contribution, therefore we stick to the original $STU$ parameters which collect the new physics contributions to the vector boson self-energies truncated at the leading order of $p^2$.}. The $W$ boson mass can then be expressed as a function of these oblique parameters~\cite{Sirlin:2012mh, Ciuchini:2013pca}
\begin{align}
    m_W=m_{W}^{\rm SM}\left[1-\frac{{\alpha}}{4\left({c_W}^{2}-{s_W}^{2}\right)}S+\frac{\alpha {c_W}^{2}}{2\left({c_W}^{2}-{s_W}^{2}\right)}T+\frac{{\alpha}}{8 {s_W}^{2}}U\right]
\label{eq:mw_STU}
\end{align}
Though new physics generally could induce further effects, the scalar extensions we considered belongs to ``universal'' theories especially that our scalars are inert without couplings to fermions.

The contribution of a general scalar multiplet of weak-isospin $j$ and hypercharge $Y$ to the oblique parameters has been calculated~\cite{Lavoura:1993nq, Zhang:2006vt}, whose expressions are presented here
\begin{equation}\label{eq:NMulti_STU}
\begin{split}
    S&=-\frac{Y}{6\pi}\sum_{l=-j}^j l\ln{m_l^2},\\
    T&=\frac{1}{4\pi s_w^2 c_w^2 m_Z^2}\left[\sum_{l=-j}^j(j^2+j-l^2)m_l^2\ln{m_l^2}-\sum_{l=-j}^{j-1}(j-l)(j+l+1)f_2(m_l,m_{l+1})\right],\\
    U&=\frac{1}{4\pi s_w^2 c_w^2 m_Z^2}\left[\sum_{l=-j}^{j-1}(j-l)(j+l+1)f_1(m_l,m_{l+1})-\sum_{l=-j}^j\frac{l^2}{3}\ln{m_l^2}\right].
\end{split}
\end{equation}
where the functions $f_1(m_1, m_2)$ and $f_2(m_1, m_2)$ are defined by
\begin{align}
    f_1(m_1, m_2)&=\int_0^1 dx x(1-x)\ln{\left[x m_1^2+(1-x)m_2^2\right]}, \\
    f_2(m_1, m_2)&=\int_0^1 dx \left[x m_1^2+(1-x)m_2^2\right]\ln{\left[x m_1^2+(1-x)m_2^2\right]}
\end{align}

To obtain the favored parameter space from considering the $W$ mass anomaly, we adopt a $\chi^2$ analysis that is defined as
\eqal{\chi^2=\chi^2(S,T,U) = \chi^2(S(g_i,m_i),T(g_i,m_i),U(g_i,m_i)),}
where the first equality is obtained from a global fit on the oblique parameters, for which we use the results in the conservative scenario of Ref.~\cite{deBlas:2022hdk} when the CDF II data is included in the fit, and the results in Ref.~\cite{Workman:2022ynf} when it is not. For the second equality, $g_i$'s and $m_i$'s are some parameters of some generic UV model, within which the oblique parameters can be expressed as functions of them as shown in Eq.~\eqref{eq:NMulti_STU}. We plot $\Delta\chi^2\equiv\chi^2-\chi^2_{\rm min}$ with $\chi^2_{\rm min}$ the minimum of $\chi^2(g_i,m_i)$ to get the allowed parameter space at 90\% and the 99\% confidence levels. 

Before going to the results, we want to comment on the construction of the $\chi^2$. To obtain the parameter space that is consistent with the CDF II measurement, one might choose to compute the $\chi^2$ from two oblique parameters $\chi^2(S(g_i,m_i), T(g_i,m_i))$, or even a single oblique parameter, $\chi^2(S(g_i,m_i))$ for example, for simplicity even when $T/U$ is non-vanishing from the prediction of certain model. However, due to the strong correlations between the oblique parameters, this procedure could lead to results that may be too optimistic such that one may then incorrectly conclude the exclusion of that model up to certain scale. Further note that a full $STU$ parameter fit including the CDF II data results in a sizable non-vanishing $U$ parameter~\cite{deBlas:2022hdk}, which indicates some new large multiplet with sufficient low masses of the components beyond the SM~\cite{Lavoura:1993nq}. Though the operator analysis within the framework of SMEFT predicts the $U$ parameter is generated by $dim$-8 operators, which is suppressed with respect to the $dim$-6 operator generated $S$ and $T$, explicit calculations of the UV models show that the values of $S$ and $T$ can be much smaller than the one of $U$ in some regions of parameter space due to accidental cancellation between different terms. Therefore we believe a full $STU$ analysis taking the correlations account should use the information as more as possible and give the best constraints on the UV models.

\begin{figure}[!htp]
\centering{
\begin{adjustbox}{max width = \textwidth}
\begin{tabular}{ccc}
\includegraphics[width=0.47\textwidth]{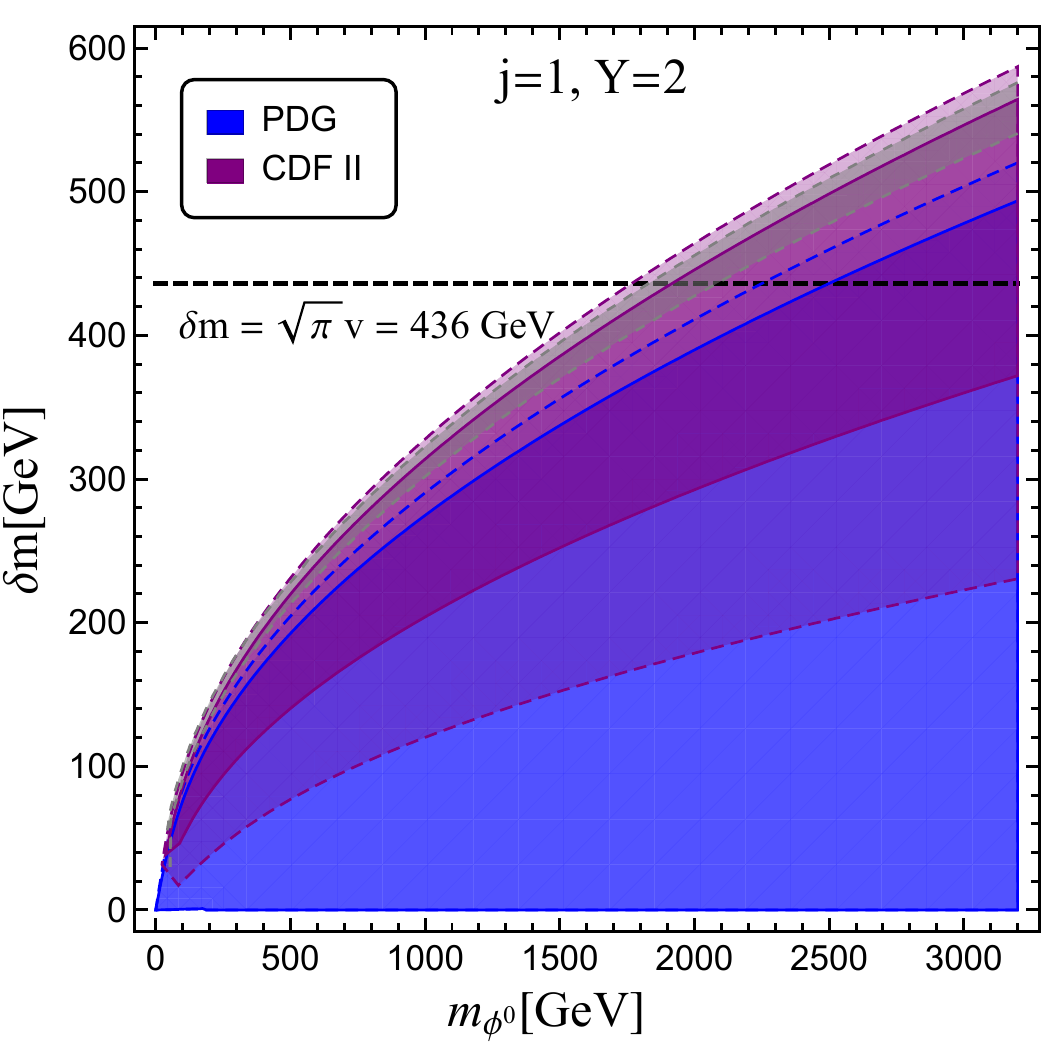} & \includegraphics[width=0.47\textwidth]{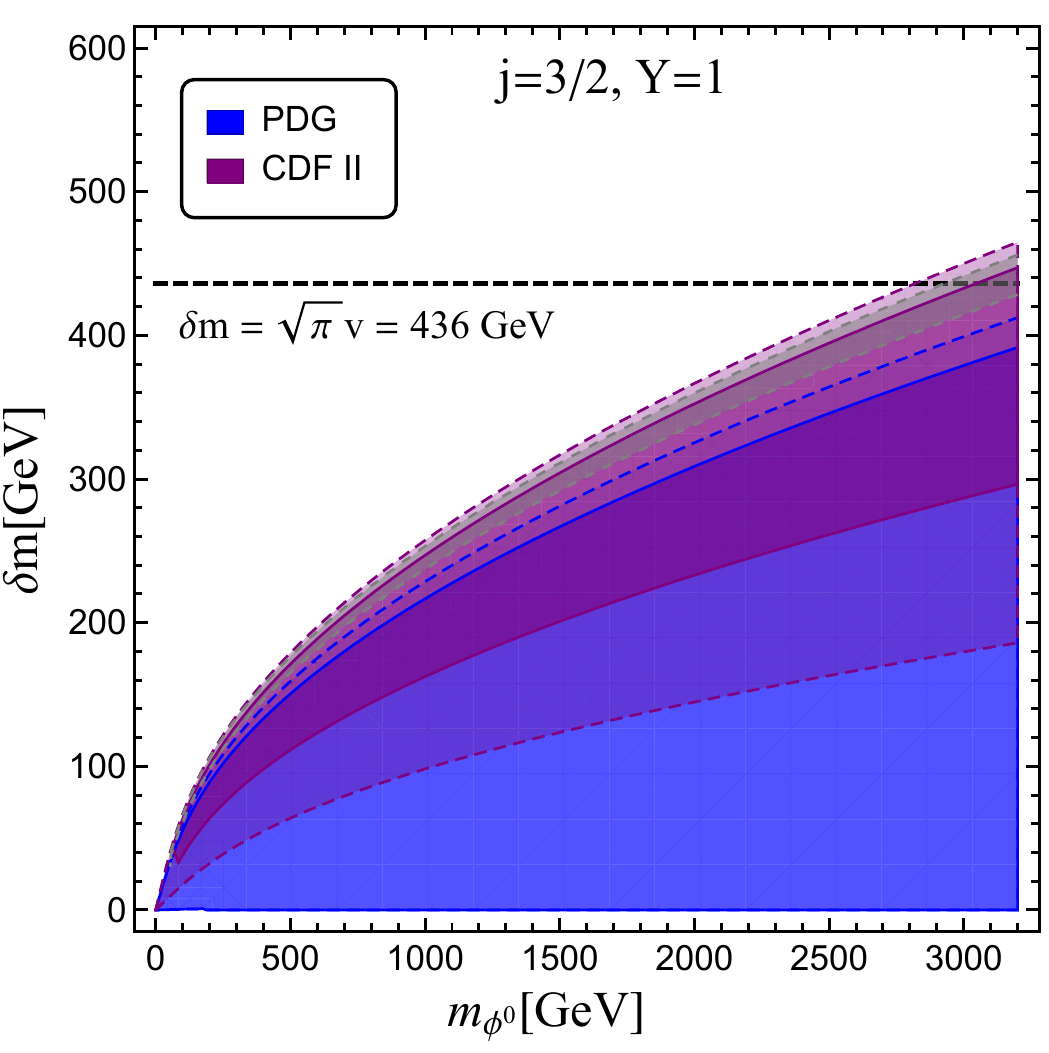}\\
\includegraphics[width=0.47\textwidth]{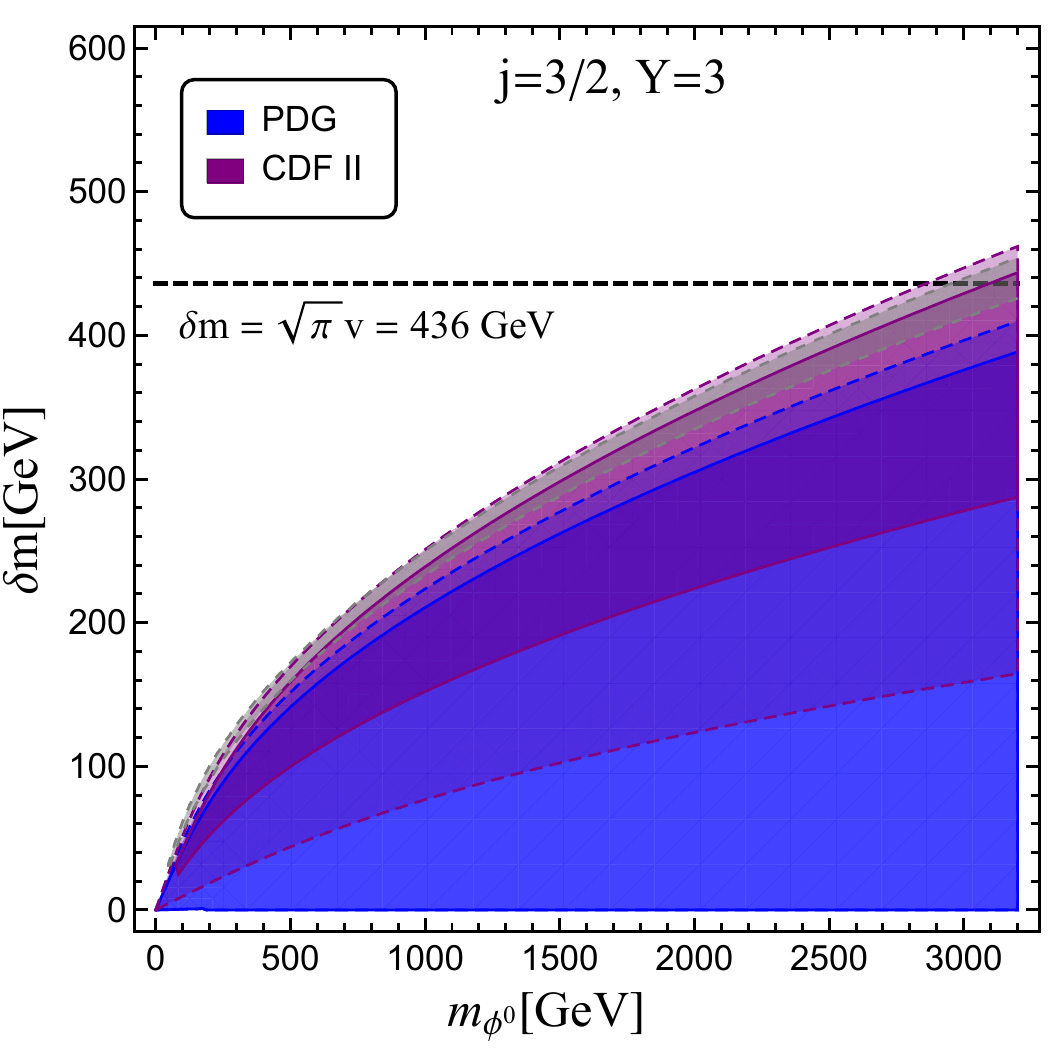} & \includegraphics[width=0.47\textwidth]{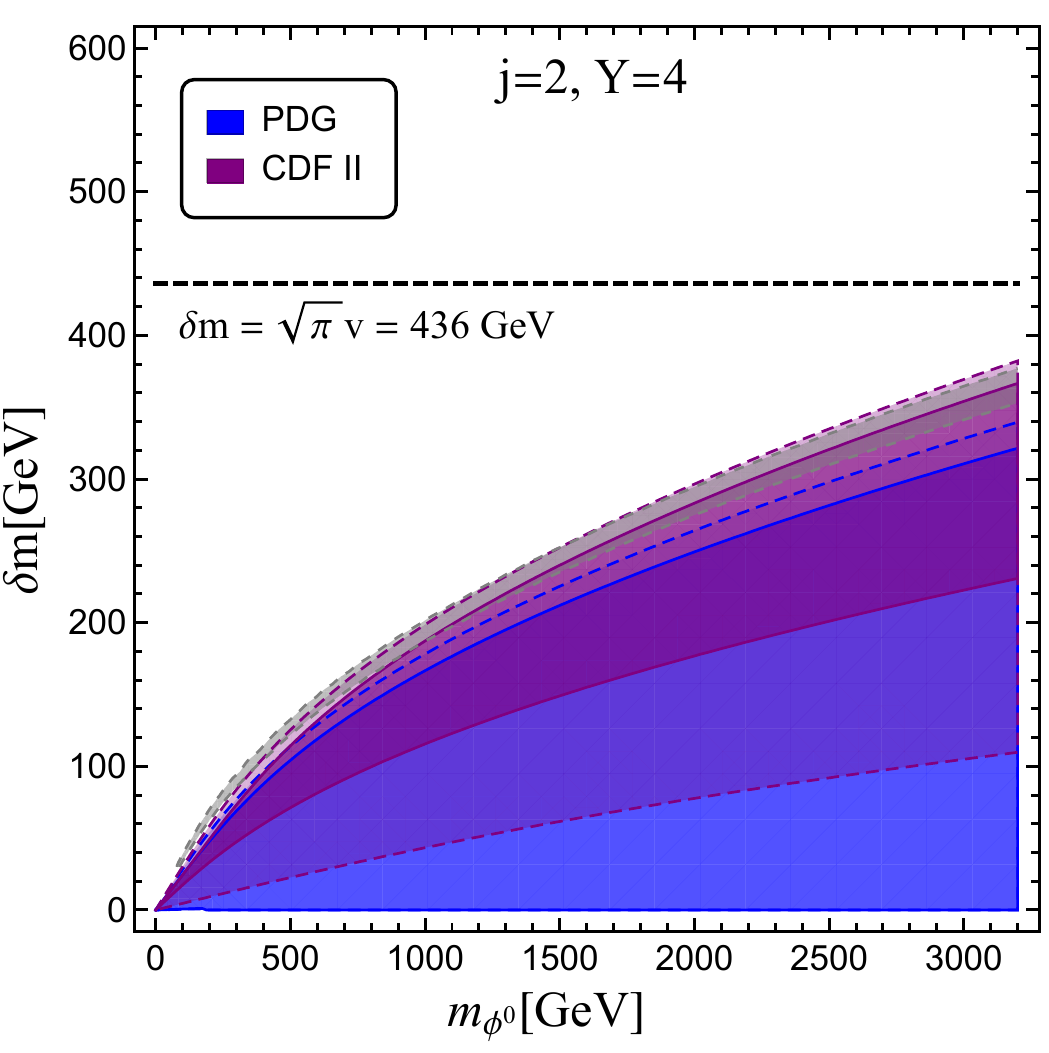}
\end{tabular}
\end{adjustbox}
}
\caption{The 90\% (solid) and the 99\% (light) CL regions for scalar multipets with quantum number $j=1, Y=2$ (upper left), $j=3/2, Y=1$ (upper right), $j=3/2, Y=3$ (lower left) and $j=2, Y=4$ (lower right) in light of the CDF II result~\cite{CDF:2022hxs}. In each plot, ``PDG'' (``CDF II'') corresponds to the constraints obtained without (with) the inclusion of the CDF II data. The gray region shows the 1$\sigma$ region consistent with the CDF-II $W$ mass measurement.}\label{plot:N-plet}
\end{figure}

The favored parameter spaces for different quantum numbers of $N$-plet are shown in Fig.~\ref{plot:N-plet},
where we define the mass splitting parameter $\delta m\equiv(-\lambda_2 v^2/4)^{1/2}$. In each plot, we use the blue (ligh-blue) color for the fit without including the CDF II data at 90\% (99\%) CL, and the purple (light-purple) for that with the inclusion of the CDF II data. The $m_W$ anomaly indicates a significant mass splitting in the spectrum while the original PDG data prefers degenerate multiplet states. Though larger hypercharge and more states could reduce the values of mass splitting to fulfill the new $STU$ parameters fit, it becomes difficult to obtain a large enough correction to $m_W$ there especially for a lighter $m_{\phi^0}$ as shown in the lower right panel. The unitarity or perturbability of the theory gives an upper bound on the mass splitting $\delta m$ shown as grey dashed line, which excludes heavier neutral states with mass above $1.7$ TeV for $j=1, Y=2$ multiplet, $2.8$ TeV for $j=3/2, Y=1$ and $2.9$ TeV for $j=3/2, Y=3$.

\subsection{\textbf{Case B}: The real and the complex triplet models with non-vanishing VEV~\label{sec:Triplets}}
We discuss in this section the general real and the complex triplet models. The real triplet model is phenomenologically interesting as it can provide a dark matter candidate and also predict the existence of a long-lived particle. See Ref.~\cite{Chiang:2020rcv} for a recent study on this point at colliders. On the other hand, the complex triplet model contains a doubly-charged Higgs particle that can decay into a same-sign dilepton final state. This channel would be very clean at colliders and serves as the smoking-gun signature for a model discovery. See Ref.~\cite{Du:2018eaw} for a detailed study at a future 100~TeV $pp$ collider and Ref.~\cite{Zhou:2022mlz} for the gravitational wave production within this model.

In light of the recent $W$ mass anomaly from CDF II experiment\cite{CDF:2022hxs}, these two triplet models are particularly interesting since relatively light real triplet and complex triplet particles are still allowed. Specifically, from the disappearing charged track searches, Ref.~\cite{Chiang:2020rcv} found that a real triplet around 250~GeV is still allowed from current data. Similarly, with a relatively large vacuum expectation value for the complex triplet model, a very light complex triplet in the $\mathcal{O}(100)$ GeV range is not ruled out yet. Such light states can generate non-negligible values of $U$ parameter. And as we pointed out ahead, non-zero VEVs of their neutral components can be developed resulting tree-level custodial symmetry breaking, which exhibits very different properties. It is these facts that motivate 
our study on these two models in explaining the the $W$ mass anomaly and exploring their effects on EWPOs.

For the following discussion, we adopt the conventions established in Ref.~\cite{Chiang:2020rcv} and Ref.~\cite{FileviezPerez:2008bj} for the complex and the real triplet models, respectively. For references, we show below the model definition for these two models:
\begin{itemize}
\item The real triplet model: The scalar sector Lagrangian of the real triplet model can be expressed as
\begin{align}
{\cal L}_{\rm real}\supset \left(D_{\mu}H \right)^{\dagger}\left(D^\mu H\right)+\rm{Tr}\left[\left(D_{\mu}\Sigma \right)^{\dagger}\left(D^{\mu}\Sigma\right)\right]-V\left(H,\Sigma \right),
\end{align}
where the SM Higgs doublet $H$ and real triplet scalar $\Sigma$ are given by
\begin{align}
H=
\begin{pmatrix}
G^+\\
\frac{1}{\sqrt{2}}\left(v_H+h+iG^0\right)
\end{pmatrix},\hspace{1cm}
\Sigma=\frac{1}{2}
\begin{pmatrix}
\Sigma^0 + v_\Sigma & \sqrt{2}\Sigma^+\\
\sqrt{2}\Sigma^- & -\Sigma^0 + v_\Sigma
\end{pmatrix},
\end{align}
with the Higgs vacuum expectation values (VEVs) $v_H^2+4v_\Sigma^2\simeq 246~\text{GeV}^2$ as determined from the muon lifetime.
$D_{\mu}\Sigma\equiv\partial_{\mu}\Sigma+ig_2\left[W_{\mu},\Sigma \right]$, with $W_{\mu}=W^a_{\mu}\tau^a/2$ and $\tau^a$ the Pauli matrices. The definition for $D_\mu H$ is standard, and the scalar potential can be written in a compact form as~\cite{FileviezPerez:2008bj}
\begin{align}
V\left(H, \Sigma \right)=-\mu^2H^{\dagger}H+\lambda_0\left(H^{\dagger}H \right)^2-\frac{1}{2}\mu^2_{\Sigma} F+\frac{b_4}{4}F^2+\frac{a_2}{2}H^{\dagger}H F + a_3 H^\dagger\Sigma H,
\end{align}
where $F\equiv\left(\Sigma^0 \right)^2+2\Sigma^+\Sigma^-$. Note that the last term above would imply the decay of $\Sigma$ into SM particles. Thus, besides a vanishing triplet VEV, an extra discrete $\mathbb{Z}_2$ symmetry needs to be applied to render the neutral component of $\Sigma$ stable as a possible dark matter candidate. On the one hand, with a vanishing triplet VEV, the charged component will receive radiative corrections at higher loops such that it will be about 166~MeV heavier than the neutral component~\cite{Cirelli:2005uq,Ibe:2012sx}. As a consequence, the charged component is long-lived and a disappearing charge track signature could be observed at colliders. This feature has been utilized in Ref.~\cite{Chiang:2020rcv} in detail and it was found that a real triplet as light as $\sim$250~GeV was still experimentally allowed. On the other hand, to saturate the dark matter relic density, it is well-known that the real triplet needs to be around 3~TeV in the minimal scenario with $a_2$=0~\cite{Cirelli:2005uq,Cirelli:2007xd}. As we will see below, for the vanishing real triplet VEV scenario, the real triplet model would be ruled out beyond 90\% CL from the CDF II data when the triplet is below $\sim$40~TeV. Thus, for the following discussion, we mainly assume the triplet to have a non-vanishing VEV such that the masses of the charged and the neutral components of $\Sigma$ will be independent. The Peskin-Takeuchi parameters in this model, to leading order in $v_\Sigma/v_H$, are given as~\cite{Forshaw:2001xq}
\begin{equation}\label{eq:RTriplet_STU}
\begin{split}
    S \approx &0,\\
    T=&\frac{4v_\Sigma^2}{\alpha v_H^2}+\frac{1}{8\pi s_{w}^{2} c_{w}^{2}m_{Z}^{2}}\left(m_{\Sigma^0}^{2}+m_{\Sigma^\pm}^{2}-\frac{2 m_{\Sigma^0}^{2} m_{\Sigma^\pm}^{2}}{m_{\Sigma^0}^{2}-m_{\Sigma^\pm}^{2}} \ln{\frac{m_{\Sigma^0}^{2}}{m_{\Sigma^\pm}^{2}}}\right),\\
    U=&-\frac{1}{3\pi}\left[\frac{3m_{\Sigma^\pm}^{2}m_{\Sigma^0}^{4}-m_{\Sigma^0}^{6}}{\left(m_{\Sigma^0}^{2}-m_{\Sigma^\pm}^{2}\right)^{3}}\ln{\frac{m_{\Sigma^0}^{2}}{m_{\Sigma^\pm}^{2}}}+\frac{5\left(m_{\Sigma^0}^{4}+m_{\Sigma^\pm}^{4}\right)-22 m_{\Sigma^0}^{2} m_{\Sigma^\pm}^{2}}{6\left(m_{\Sigma^0}^{2}-m_{\Sigma^\pm}^{2}\right)^{2}}\right]~,
\end{split}
\end{equation}
where the $T$ parameter is composed of two parts, tree-level CSV contribution (1st term) from non-vanishing VEV of the triplet 
and one-loop CSV contribution (2nd term) from mass splitting between charged and neutral components. The contribution 
of order $\mathcal{O}(m_Z/m_\Sigma^\pm)$
and higher are neglected.

\item The complex triplet model: We parameterize the Lagrangian of this complex triplet model as
\eqal{\mathcal{L}\supset\mathcal{L}_{\rm kinetic}(\Delta) - V(H,\Delta),}
with
\begin{align}
\mathcal{L}_{\rm kinetic}(\Delta)&={\rm{Tr}}[(D_\mu \Delta)^\dagger (D^\mu \Delta)],\\
V(H,\Delta)&= - m^2 H^\dagger H + M^2{\rm{Tr}}(\Delta^\dagger\Delta)+\left[\mu H^Ti\tau_2\Delta^\dagger H+\rm{h.c.}\right]+\lambda_1(H^\dagger H)^2 \nonumber\\
&~~~~+\lambda_2\left[\rm{Tr}(\Delta^\dagger\Delta)\right]^2 +\lambda_3\rm{Tr}[ \Delta^\dagger\Delta \Delta^\dagger\Delta]
+\lambda_4(H^\dagger H)\rm{Tr}(\Delta^\dagger\Delta)\nonumber\\
&~~~~+\lambda_5 H^\dagger\Delta\Delta^\dagger H.\label{eq:tripletpotential}
\end{align}
The covariant derivative on $\Delta$ is defined as $D_\mu \Delta\equiv\partial_\mu \Delta + {ig}[\tau^aW_\mu^a,\Delta]/{2}+{ig'Y_{\Delta}}B_\mu\Delta/{2}$ with $Y_{\Delta}=2$. The components of the SM Higgs doublet $H$ is the same as defined above, while those for the complex triplet are defined as
\eqal{
\Delta =
\left(
\begin{array}{cc}
\frac{\Delta^+}{\sqrt{2}} & H^{++}\\
\frac{1}{\sqrt{2}}(\delta+v_\Delta+i\eta) & -\frac{\Delta^+}{\sqrt{2}} 
\end{array}\right).}
The complete mass spectrum of this model can be found, for example, in Ref.~\cite{Du:2018eaw}, which we cite in the following:
\eqal{
m_h^2\simeq2v^2\lambda_1,\,m_H \simeq M_\Delta \simeq m_A,\,m_{H^\pm}^2\simeq M_\Delta^2-\frac{\lambda_5}{4}v_H^2,\, m_{H^{\pm\pm}}^2\simeq M_\Delta^2-\frac{\lambda_5}{2}v_H^2,}
taking into account the fact that $v_\Delta\ll v_H$ as required by the $\rho$ parameter. While the triplet VEV $v_\Delta$ is relatively tiny compared to $v_H$, its magnitude has a non-trivial impact on the phenomenological side. In particularly, when $v_\Delta$ is below $\sim10^{-4}$~GeV, the same-sign dilepton channel dominates the decay of the doubly charged triplet, acting as the smoking-gun signature for model discovery. In contrast, when $v_\Delta$ is above $\sim10^{-4}$~GeV but fulfills the requirement of the $\rho$ parameter, the same-sign di-$W$ boson final state would dominate instead -- See Ref.~\cite{Du:2018eaw} for the details. However, as we will see below, the CDF II data does not have any sensitivity to such small values of $v_\Delta$ ($v_\Delta\lesssim 1$~GeV), it is thus important to keep in mind the impact on the phenomenologies from $v_\Delta$. The $STU$ parameters are calculated in Refs.~\cite{Lavoura:1993nq, Chun:2012jw},
\begin{equation}\label{eq:CTriplet_STU}
\begin{split}
S=&-\frac{1}{3 \pi} \ln \frac{m_{+1}^{2}}{m_{-1}^{2}}-\frac{2}{\pi} \sum_{T_{3}=-1}^{+1}\left(T_{3}-Q s_{w}^{2}\right)^{2} \xi\left(\frac{m_{T_{3}}^{2}}{m_{Z}^{2}}, \frac{m_{T_{3}}^{2}}{m_{Z}^{2}}\right), \\
T=&-\frac{2v_\Delta^2}{\alpha\left(v_H^2+4v_\Delta^2\right)}+\frac{1}{16 \pi c_{w}^{2} s_{w}^{2}} \sum_{T_{3}=-1}^{+1}\left(2-T_{3}\left(T_{3}-1\right)\right) \eta\left(\frac{m_{T_{3}}^{2}}{m_{Z}^{2}}, \frac{m_{T_{3}-1}^{2}}{m_{Z}^{2}}\right), \\
U=&\frac{1}{6 \pi} \ln \frac{m_{0}^{4}}{m_{+1}^{2} m_{-1}^{2}}+\frac{1}{\pi} \sum_{T_{3}=-1}^{+1}\left[2\left(T_{3}-Q s_{w}^{2}\right)^{2} \xi\left(\frac{m_{T_{3}}^{2}}{m_{Z}^{2}}, \frac{m_{T_{3}}^{2}}{m_{Z}^{2}}\right)\right.\\
&\left.-\left(2-T_{3}\left(T_{3}-1\right)\right) \xi\left(\frac{m_{T_{3}}^{2}}{m_{W}^{2}}, \frac{m_{T_{3}-1}^{2}}{m_{W}^{2}}\right)\right],
\end{split}
\end{equation}
with
\eqal{
m_{+1,0,-1}=&m_{H^{\pm\pm},H^\pm,H}\\
\xi(x, y)=& \frac{4}{9}-\frac{5}{12}(x+y)+\frac{1}{6}(x-y)^{2} \\
&+\frac{1}{4}\left[x^{2}-y^{2}-\frac{1}{3}(x-y)^{3}-\frac{x^{2}+y^{2}}{x-y}\right] \ln \frac{x}{y}-\frac{1}{12} d(x, y) f(x, y) \\
d(x, y)=&-1+2(x+y)-(x-y)^{2} \\
f(x, y)=&\left\{\begin{array}{ll}
-2 \sqrt{d(x, y)}\left[\arctan \frac{x-y+1}{\sqrt{d(x, y)}}-\arctan \frac{x-y-1}{\sqrt{d(x, y)}}\right], & \text { for } \quad d(x, y)>0 \\
\sqrt{-d(x, y)} \ln \left[\frac{x+y-1+\sqrt{-d(x, y)}}{x+y-1-\sqrt{-d(x, y)}}\right], & \text { for } \quad d(x, y) \leq 0
\end{array}\right.\\
\eta(x, y)=& x+y-\frac{2 x y}{x-y} \ln \frac{x}{y}.
}
\end{itemize}

\begin{figure}[!htp]
\centering{
\begin{adjustbox}{max width = \textwidth}
\begin{tabular}{ccc}
\includegraphics[width=0.47\textwidth]{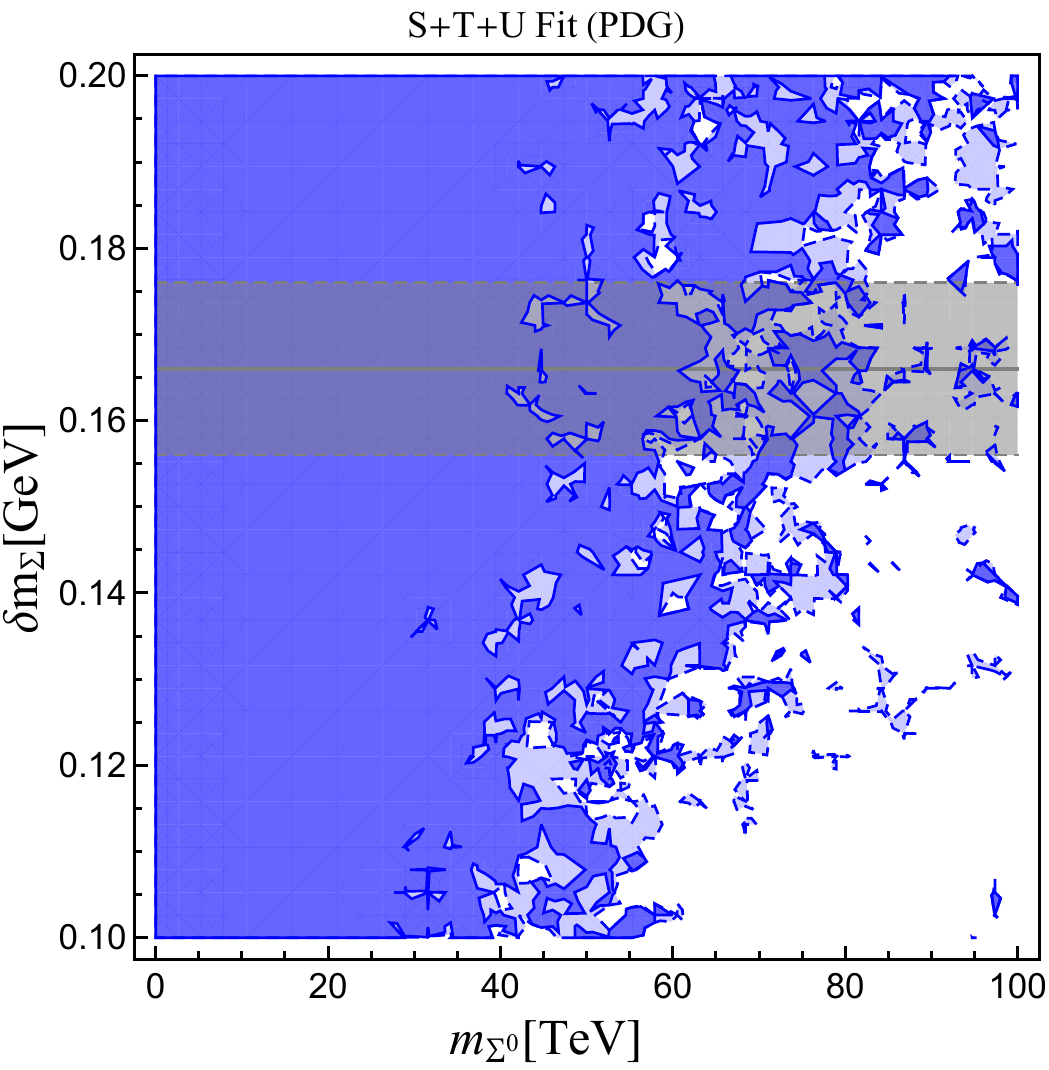} & \includegraphics[width=0.47\textwidth]{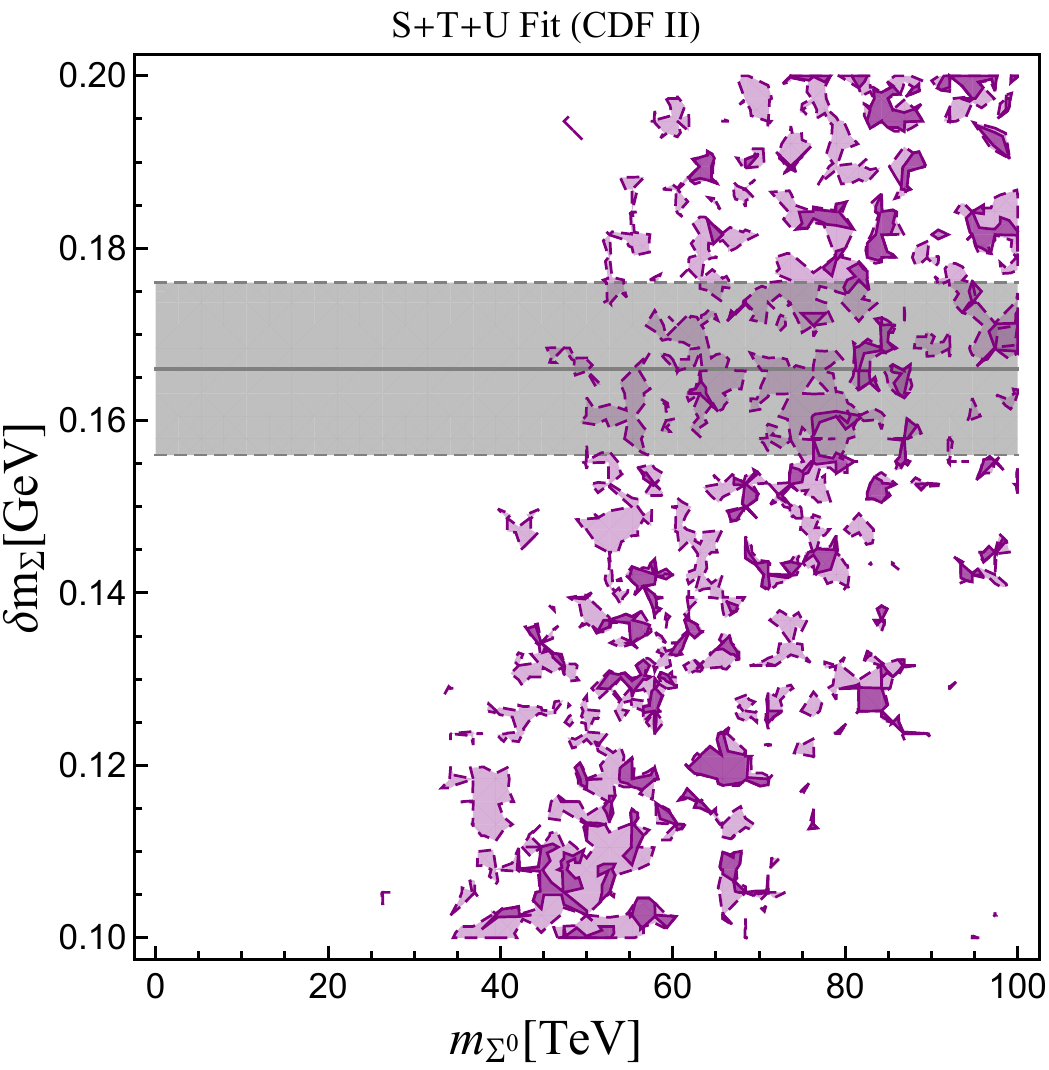}\\
\includegraphics[width=0.47\textwidth]{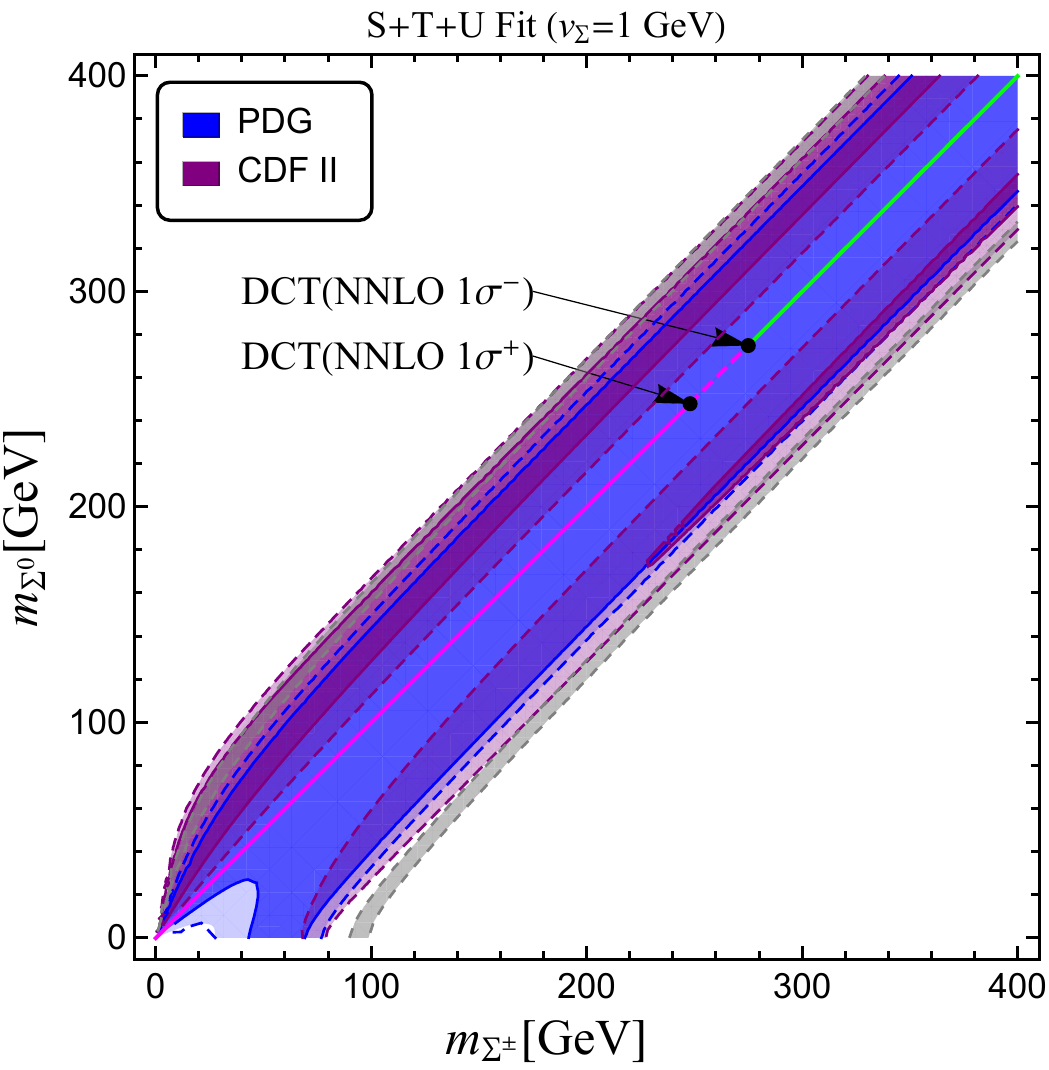} & \includegraphics[width=0.47\textwidth]{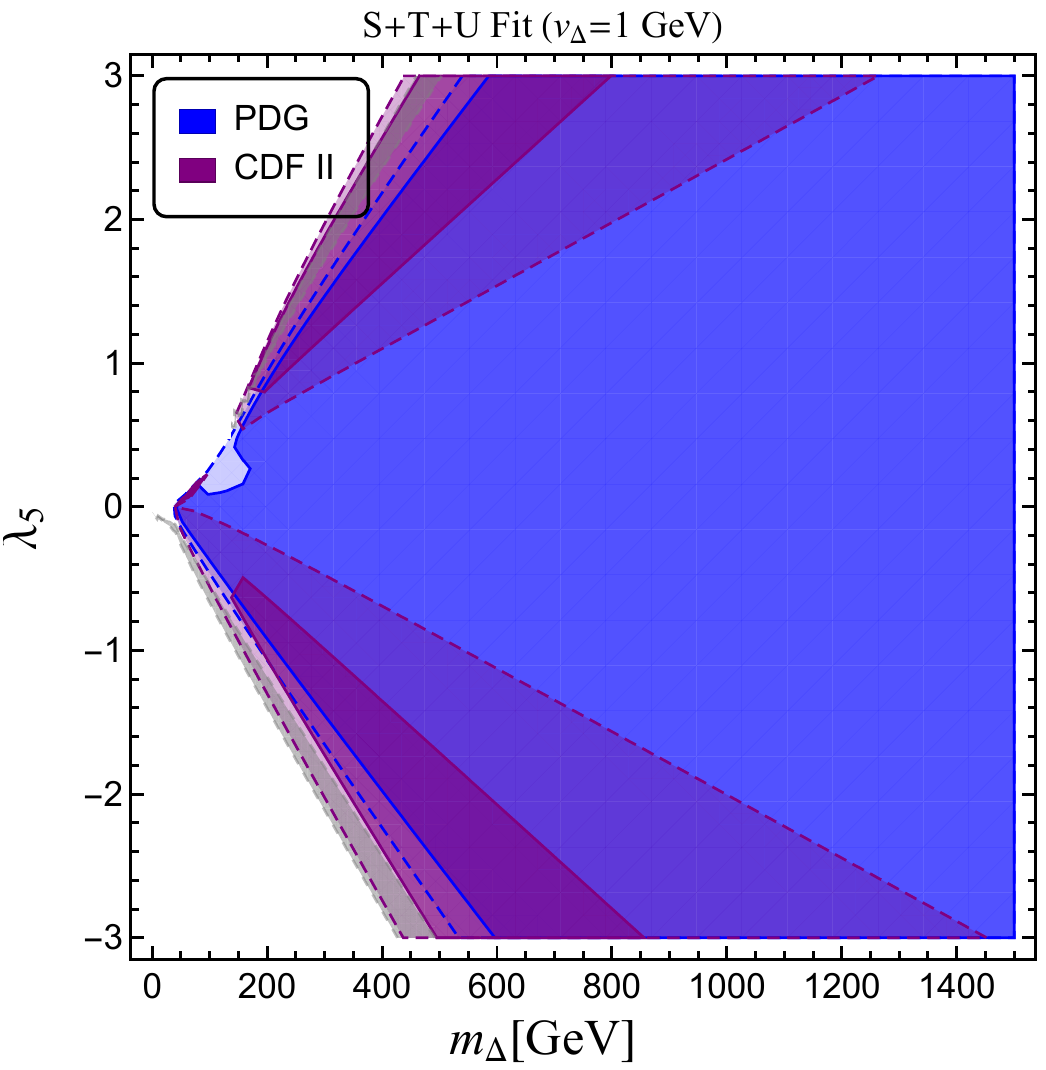}
\end{tabular}
\end{adjustbox}
}
\caption{The 90\% (solid) and the 99\% (light) CL regions in for the real (upper row + lower left panel) and the complex (lower right) triplet models in light of the CDF II result~\cite{CDF:2022hxs}. See text for details.}\label{plot:realandcomplex}
\end{figure}

Using the full $\chi^2$ obtained by taking the correlations among the oblique parameters into account, we present the results for the real and the complex triplet models in Fig.~\ref{plot:realandcomplex} with same color notation as in Fig.~\ref{plot:N-plet}.
Several points merit addressing:
\begin{itemize}
\item The upper row of Fig.~\ref{plot:realandcomplex} corresponds to the real triplet model in the vanishing VEV scenario, and the lower left plot of Fig.~\ref{plot:realandcomplex} to that in the non-vanishing VEV scenario. Note that in the former case, as discussed above, the mass splitting between the charged and neutral particles in the triplet is fixed at about 166~MeV at one-loop order and will be further modified by a few MeV at two-loop level. This is reflected by the gray region in the first row. From these two plots, it is clear that the CDF II data will generically prefer a heavier triplet. On the other hand, in the non-vanishing VEV scenario, masses of the two triplet particles can vary independently. Though the CDF-II data indicates manifest custodial symmetry breaking at some level which prefers a non-vanishing VEV, as the low energy measurement of $\rho$ parameter (the ratio of neutral-to-charged current interaction strengths) is constrained to be close to one~\cite{ParticleDataGroup:2020ssz}, the value of $v_\Sigma$ can not exceed few GeV, indicating that $v_\Sigma\ll v$ is a good approximation. Therefore we only show the result at a fixed value of $v_\Sigma$ ($v_\Sigma=1~$GeV) by performing a two-parameter fit instead of a full three-parameter fit. The allowed parameter space shrinks when we increase $v_\Sigma$ unless $v_\Sigma$ is below $\mathcal{O}(\text{GeV})$ where the sensitivity of our $\chi^2$ to $v_\Sigma$ is lost. We find that, with or without the inclusion of the CDF II data, the preferred parameter space lies along the diagonal region on the $M_{\Sigma^0}-M_{\Sigma^\pm}$ plane with the difference that the CDF II data prefers generically a larger mass splitting between the two triplet particles. For reference, we also show the constraints from the disappearing charged track search obtained in Ref.~\cite{Chiang:2020rcv} when the two particles have degenerated mass spectrum in magenta, where the upper (lower) point corresponds to a mass splitting of 172 (160)~MeV. The green color indicates the still allowed parameter space from the current disappearing charged track search. 
\item We also comment on the appearance of separated islands in the first row of figure\,\ref{plot:realandcomplex} at large $M_{\Sigma^0}$'s. These islands are merely a reflection of partial cancellation between different types of corrections to the oblique parameters. For illustration, one can look at the $\delta\chi^2$ in this vanishing VEV scenario from PDG, for which we find\footnote{This formula here is from numerical calculation for illustration, and both $\delta M_\Sigma$ and $M_{\Sigma^0}$ should be understood as dimensionless numbers of their corresponding values measured in GeV.}
\eqal{(\delta\chi^2)^{\rm PDG}\simeq &\, 2.4774 + \frac{8.96 \delta M_\Sigma}{M_{\Sigma^0}} + \frac{3.18\times 10^{-2} \delta M_\Sigma^2}{M_{\Sigma^0}^2} \nonumber \\
&\, + 1.82\times 10^{-6} \delta M_\Sigma^4 + \frac{5.96\times 10^{-3} \delta M_\Sigma^4}{M_{\Sigma^0}^2} + \frac{13.24 \delta M_\Sigma^4}{M_{\Sigma^0}^4}\nonumber\\
&\, - 4.84\times10^{-3} \delta M_\Sigma^2 - \frac{6.34\times 10^{-3} \delta M_\Sigma^3}{M_{\Sigma^0}} - \frac{8.00 \delta M_\Sigma^3}{M_{\Sigma^0}^3}.}
Clearly, accidental partial cancellation between the third line and the first two lines is expected for different values of $\delta M_\Sigma$ and $M_{\Sigma^0}$. For this reason, increasing or decreasing the numerical precision to obtain these plots will lead to slightly different results, thus our results in the first row of figure\,\ref{plot:realandcomplex} are only for quantitative illustration instead of trying to be exact. However, we stress that the pattern of these plots, and thus our conclusion presented above and below, remains unchanged.
\item Results for the complex triplet model is shown in the lower right panel of Fig.~\ref{plot:realandcomplex}. Similar to the real triplet case, we find that $v_\Delta$ can not exceed few GeV. Note that the non-vanishing $v_\Delta$ gives an incorrect sign contribution to explain the CDF-II results, we expect that the 1-loop correction dominates over or at least at the same level as the tree-level correction, which also indicates approximation $v_\Delta\ll v$ is valid. For this reason, we also fix $v_\Delta$ to 1~GeV for illustration, and increasing the value of VEV of the complex triplet shrinks the allowed parameter region unless $v_\Delta\lesssim 1$~GeV where our $\chi^2$ is insensitive to $v_\Delta$ any more. Clearly, the CDF II data implies a much more constrained parameter space than the previous experiments. Especially, with $\lambda_5$  within the perturbativity bound, a light triplet below about 1~TeV would be able to explain the $W$ mass anomaly. One could thus expect to exploit the same-sign di-lepton (di-$W$) channel for the model discovery at future colliders~\cite{Du:2018eaw}.
\item The contributions to the $T$ parameter from the non-zero VEVs in the real and complex triplet models have opposite signs. The non-vanishing $v_\Delta$ of complex triplet partially cancels the radiative corrections in the $T$ parameter which might result in rather small $T$ values compared to $U$, while the effect of the VEV of a real triplet owns exact opposite property. Therefore a complete three-parameter fit including the VEV prefers a non-vanishing VEV for the real triplet which is also favored by lifting the degeneracy between the charged and neutral components.
\end{itemize}
\section{Custodial symmetric scalar sector extensions of the SM, Georgi-Machacek model as an example \label{sec:GM}}


In general, a custodial symmetric scalar sector needs at least two extra scalar multiplets besides the original SM doublet. The simplest model of this kind is the GM model with a real triplet and a complex triplet with $Y=2$, which are combined into a bitriplet under the global $SU(2)_L\times SU(2)_R$ symmetry. One simple way to understand why this model preserves the custodial symmetry is via the view of EFT, that the Wilson coefficients of dim-6 custodial breaking operator $\mathcal{O}_{HD}$ after integrating the real and complex scalar triplet have opposite signs~\cite{Dawson:2017vgm, Corbett:2017ieo}, which could cancel each other by requiring triplets to form a bi-triplet.

To generalize this model, it is useful to consider large bi-multiplet representations of $SU(2)_L\times SU(2)_R$ symmetry. After taking into account unitarity constraint, there are only four models including the GM one~\cite{Logan:2015xpa, Logan:2015fka}, which are listed in Tab.~\ref{tab:GMM}
\begin{table}[!htp]
\centering
\begin{tabular}{ccccc}
\hline\hline
Model name & $SU(2)_L\times SU(2)_R$ reps. & $j$ & $Y$ & real/complex \\
\hline\hline
\multirow{3}{*}{GM} & \multirow{3}{*}{$(2\times 2)+(3\times 3)$} & $1/2$ & $1$ & $\mathbb{C}$ \\
  &  & $1$ & $2$ & $\mathbb{C}$ \\
  &  & $1$ & $0$ & $\mathbb{R}$ \\
\hline
\multirow{3}{*}{GGM4} & \multirow{3}{*}{$(2\times 2)+(4\times 4)$} & $1/2$ & $1$ & $\mathbb{C}$ \\
  &  & $3/2$ & $3$ & $\mathbb{C}$ \\
  &  & $3/2$ & $1$ & $\mathbb{C}$ \\
\hline
\multirow{4}{*}{GGM5} & \multirow{4}{*}{$(2\times 2)+(5\times 5)$} & $1/2$ & $1$ & $\mathbb{C}$ \\
  &  & $2$ & $4$ & $\mathbb{C}$ \\
  &  & $2$ & $2$ & $\mathbb{C}$ \\
  &  & $2$ & $0$ & $\mathbb{R}$ \\
\hline
\multirow{4}{*}{GGM6} & \multirow{4}{*}{$(2\times 2)+(6\times 6)$} & $1/2$ & $1$ & $\mathbb{C}$ \\
  &  & $5/2$ & $5$ & $\mathbb{C}$ \\
  &  & $5/2$ & $3$ & $\mathbb{C}$ \\
  &  & $5/2$ & $1$ & $\mathbb{C}$ \\
\hline
\end{tabular}
\caption{Scalar field contents in the generalized Georgi–Machacek models.}\label{tab:GMM}
\end{table}

All of these models share the unique features. Since the GM model is the simplest one and is the least constrained,
in this section, we use it as an example to discuss whether these kind of models can solve the $m_W$ anomaly and how the VEVs and custodial symmetry affect the $W$ mass and EWPOs. The custodial symmetry could be preserved at the tree-level even when the VEVs of the two triplets are set to be same. Thus the VEVs of the triplets could be as large as $\mathcal{O}(\rm{GeV})$.

As the custodial symmetry originate from $SU(2)_L\times SU(2)_R$
before electroweak symmetry breaking~(EWSB), the 
Lagrangian in GM model preserves $SU(2)_L\times SU(2)_R$ symmetry. It could be explicitly shown when the scalars are 
written in bidoublet and bitriplet forms, which are
\begin{align}
    \Phi =  \left(
    i\tau^2 H^*~~ H \right), \qquad
\Xi = \left( \begin{array}{ccc} \left( \Delta^0 \right)^* & \Sigma^+ & \Delta^{++} \\ - \left( \Delta^+ \right)^* & \Sigma^0 & \Delta^+ \\ \left( \Delta^{++} \right)^* & -\left( \Sigma^+ \right)^* & \Delta^0 \end{array} \right)~.
\end{align}
Where $\Phi$ is bidoublet form of the SM Higgs, $\Xi$ combines a real scalar triplet $\Sigma$ and a complex scalar triplet $\Delta$. The scalar potential 
is written as
\begin{align}
V(\Phi,\Xi) = 
& \frac12 m_{\Phi}^2 {\rm tr} \left[ \Phi^\dagger \Phi \right] + \frac12 m_{\Xi}^2 {\rm tr} \left[ \Xi^\dagger \Xi \right] + \lambda_1 \left( {\rm tr} \left[ \Phi^\dagger \Phi \right] \right)^2 + \lambda_2 \left( {\rm tr} \left[ \Xi^\dagger \Xi \right] \right)^2 \nonumber \\
& + \lambda_3 {\rm tr} \left[ \left( \Xi^\dagger \Xi \right)^2 \right] + \lambda_4 {\rm tr} \left[ \Phi^\dagger \Phi \right] {\rm tr} \left[ \Xi^\dagger \Xi \right] 
+ \lambda_5 {\rm tr} \left[ \Phi^\dagger t^a \Phi t^b \right] {\rm tr} \left[ \Xi^\dagger T^a \Xi T^b \right] \nonumber \\
& - M_1 {\rm tr} \left[ ( \Phi^\dagger t^a \Phi t^b ) ( U \Xi  U^\dagger )_{ab} \right] - M_2 {\rm tr} \left[ ( \Xi^\dagger t^a \Xi t^b ) (  U \Phi U^\dagger )_{ab} \right] ~,\label{eq:GMpotential}
\end{align}
where $t^a$, $T^a$ are SU(2) generators in $2\times 2$ and $3\times 3$ representations, and the matrix $U$ is given by
\begin{align}
    U=\begin{pmatrix}
    -\frac{1}{\sqrt{2}} & 0 & \frac{1}{\sqrt{2}} \\
    -\frac{i}{\sqrt{2}} & 0 & \frac{i}{\sqrt{2}} \\
    0 & 1 & 0
    \end{pmatrix}.
\end{align}
The vacuum expectation values of the triplets $\Sigma$ and $\Delta$ are same required by the custodial symmetry, so their difference $\nu=\langle \Sigma^0 \rangle-\langle \Delta^0 \rangle =0$. For simplicity, we define $v_\chi\equiv\braket{\Sigma^0}=\braket{\Delta^0}$, and $v_H^2+8v_\chi^2\simeq 246~\text{GeV}^2$.

In the GM model, there are 5-plet ($H_5$), 3-plet ($H_3$), and two singlet Higgs bosons under the classification of the custodial $SU(2)_V$ symmetry, where the latter two can mix with each other with mixing angles $\beta$ and $\alpha$. The mixing angle $\beta$ is defined by $t_\beta=2\sqrt{2}v_\chi/v_H$\footnote{We use $s_\theta$, $c_\theta$ and $t_\theta$ to denote $\sin\theta$, $\cos\theta$ and $\tan\theta$ respectively for some angle $\theta$.}, while $\alpha$ is determined by the quartic coupling constants in the Higgs potential, which is treated as a free input parameter in our analysis. The custodial symmetry requires mass degeneracy within each Higgs multiplet, therefore we denote $m_3$ and $m_5$ as the masses of 5-plet and 3-plet respectively. Among the mass eigenstates of the two singlet fields, one ($h$) should be identified as the observed 125 GeV SM-like Higgs, and the other one ($H$) has mass $m_H$.

The $STU$ parameters in GM model are calculated in Refs.~\cite{Kanemura:2013mc,Hartling:2014aga}, and here we adopt the results given in Ref.~\cite{Kanemura:2013mc}. It is well-known that the predicted $T$ parameter contains some UV divergence, which is a result that the custodial symmetry in the GM potential would be broken at loop level~\cite{Gunion:1990dt, Blasi:2017xmc, Keeshan:2018ypw}. Accordingly, to absorb the infinities of the $T$ parameter at the one-loop level a counterterm of $\nu$ has to be added, which however introduces some renormalization scale dependence in the results. The previous experimental values for the $T$ parameter are nearly zero~\cite{ParticleDataGroup:2020ssz}, which means the custodial symmetry is almost preserved, so that earlier studies on GM model usually set $T$ to be zero as an input parameter~\cite{Blasi:2017xmc,Keeshan:2018ypw,Chiang:2018xpl,Englert:2013zpa,Kanemura:2013mc}. Ref.~\cite{Blasi:2017xmc,Keeshan:2018ypw} estimate the amount of the custodial symmetry violation effect at higher energy scale when the $\rho$ parameter at weak scale is fixed to be almost one. Given the $m_W$ measured from CDF II, a larger $T$ value is predicted, indicating a large custodial symmetry breaking. A more self-consistent way to deal with renormalization is adding explicitly custodial symmetry breaking terms to the Lagrangian as performed in Ref.~\cite{Chiang:2018xpl}, which typically generates misaligned VEVs. Such misalignment can address the CDF II anomaly as shown in Ref.~\cite{Du:2022brr, Chen:2022ocr}. However, this procedure indeed makes the model deviate from the 
original idea of the GM model. Hence here we consider the custodial symmetric case with aligned VEVs assuming that some fine-tuning exists at the UV to align two triplet VEVs, and focus on the pure one-loop effects ($STU$ parameters) on EWPOs and $W$ mass. To accommodate the lack of knowledge of the UV, we marginalize over the $T$ parameter in our $\chi^2$ analysis.

\begin{figure}[!htp]
\centering{
\begin{adjustbox}{max width = \textwidth}
\begin{tabular}{cc}
\includegraphics[width=0.4\textwidth]{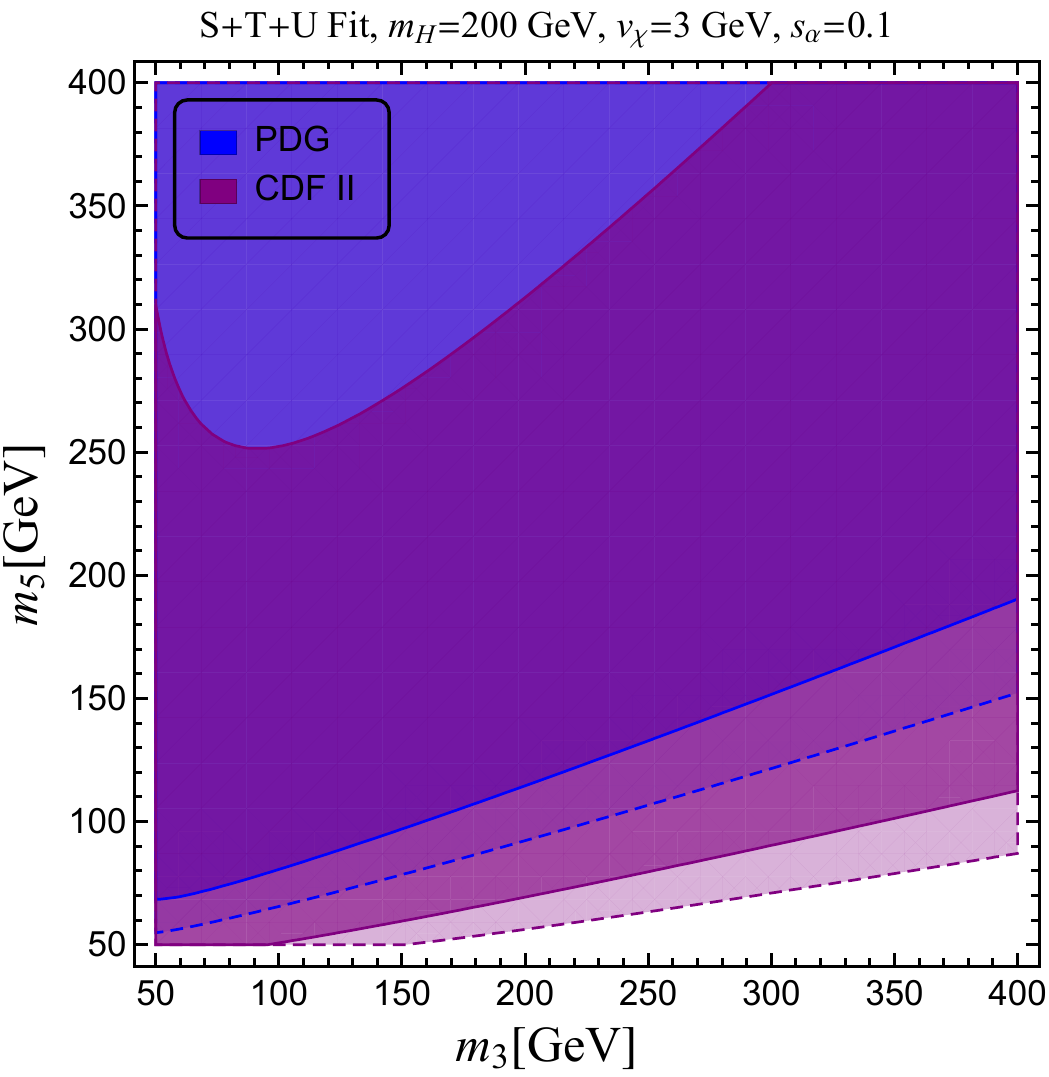} & 
\includegraphics[width=0.4\textwidth]{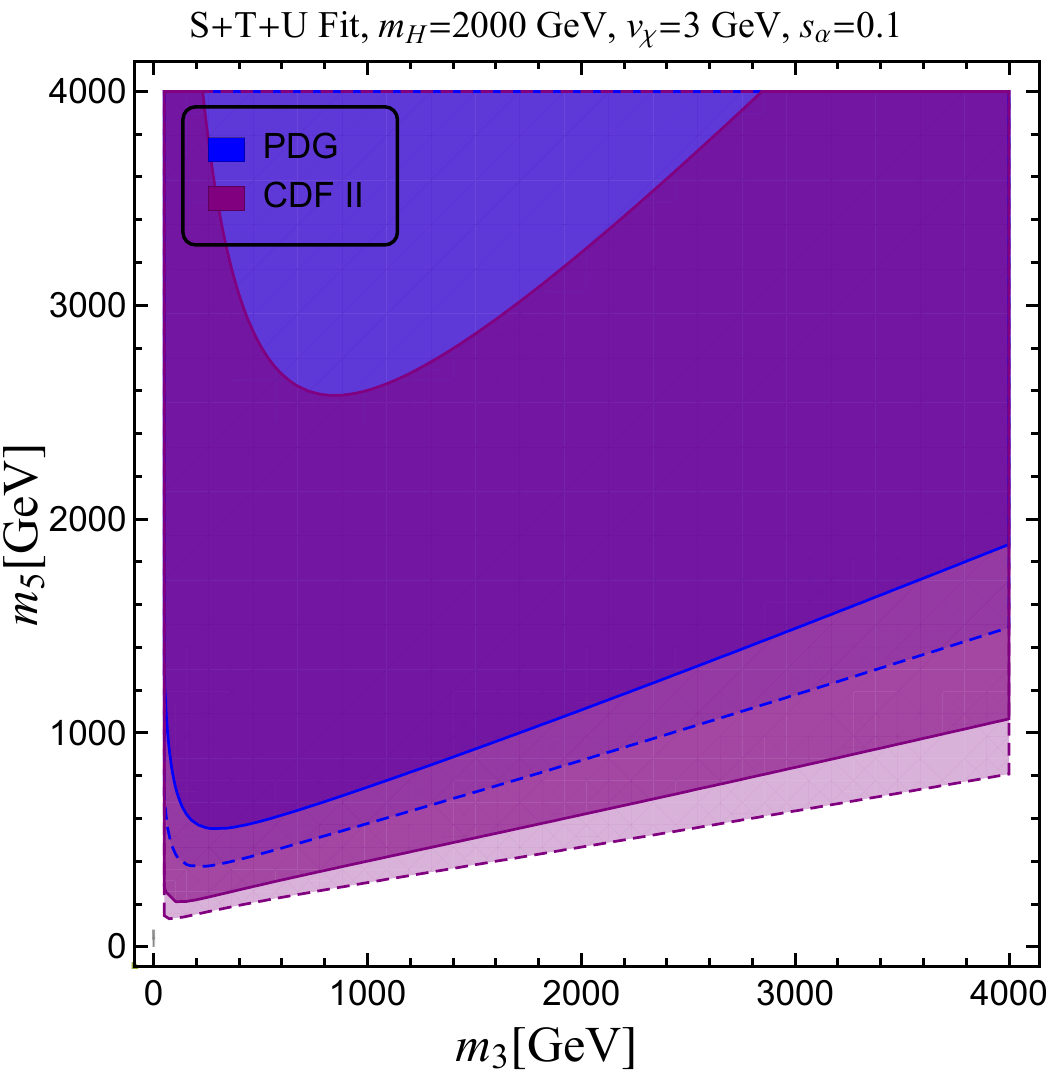} \\ 
\includegraphics[width=0.4\textwidth]{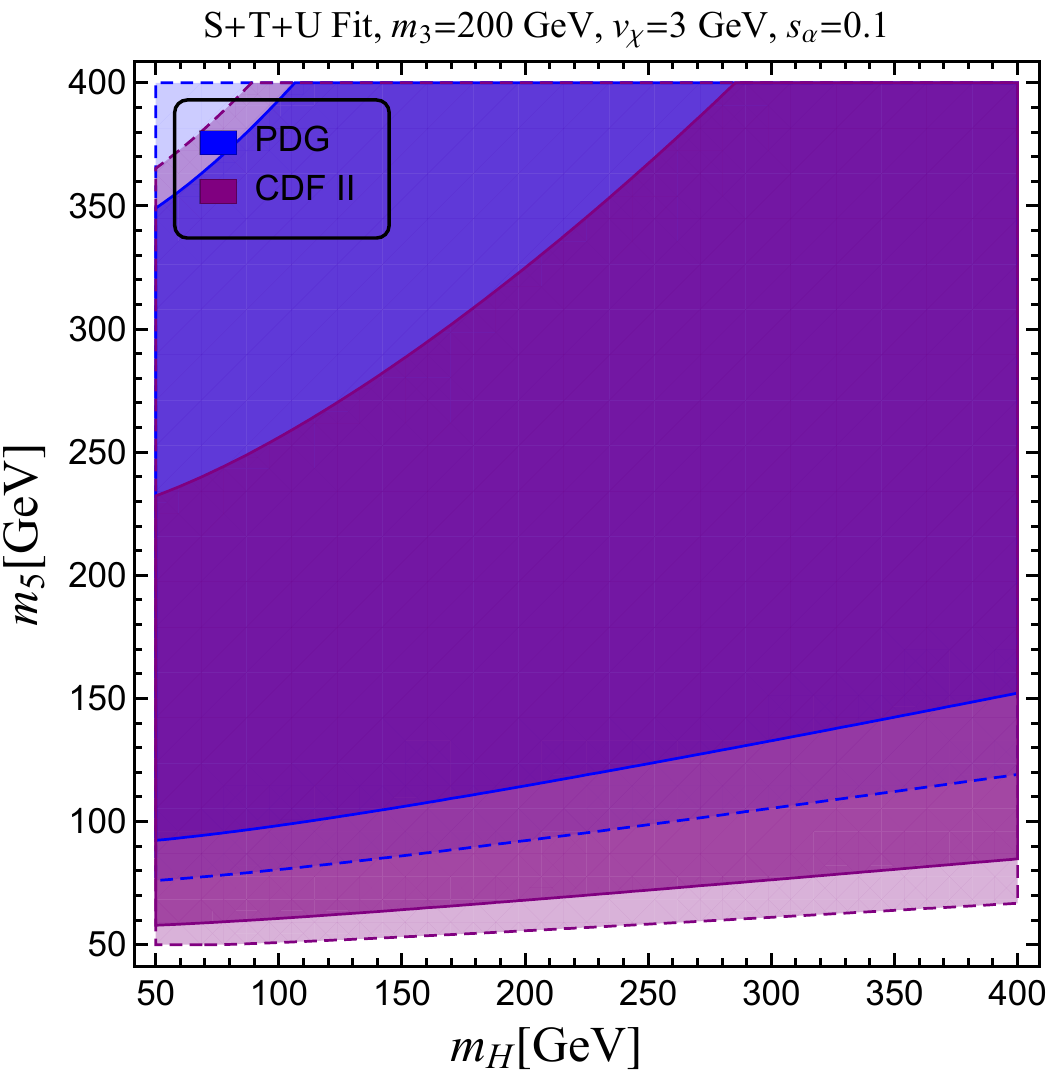} &
\includegraphics[width=0.4\textwidth]{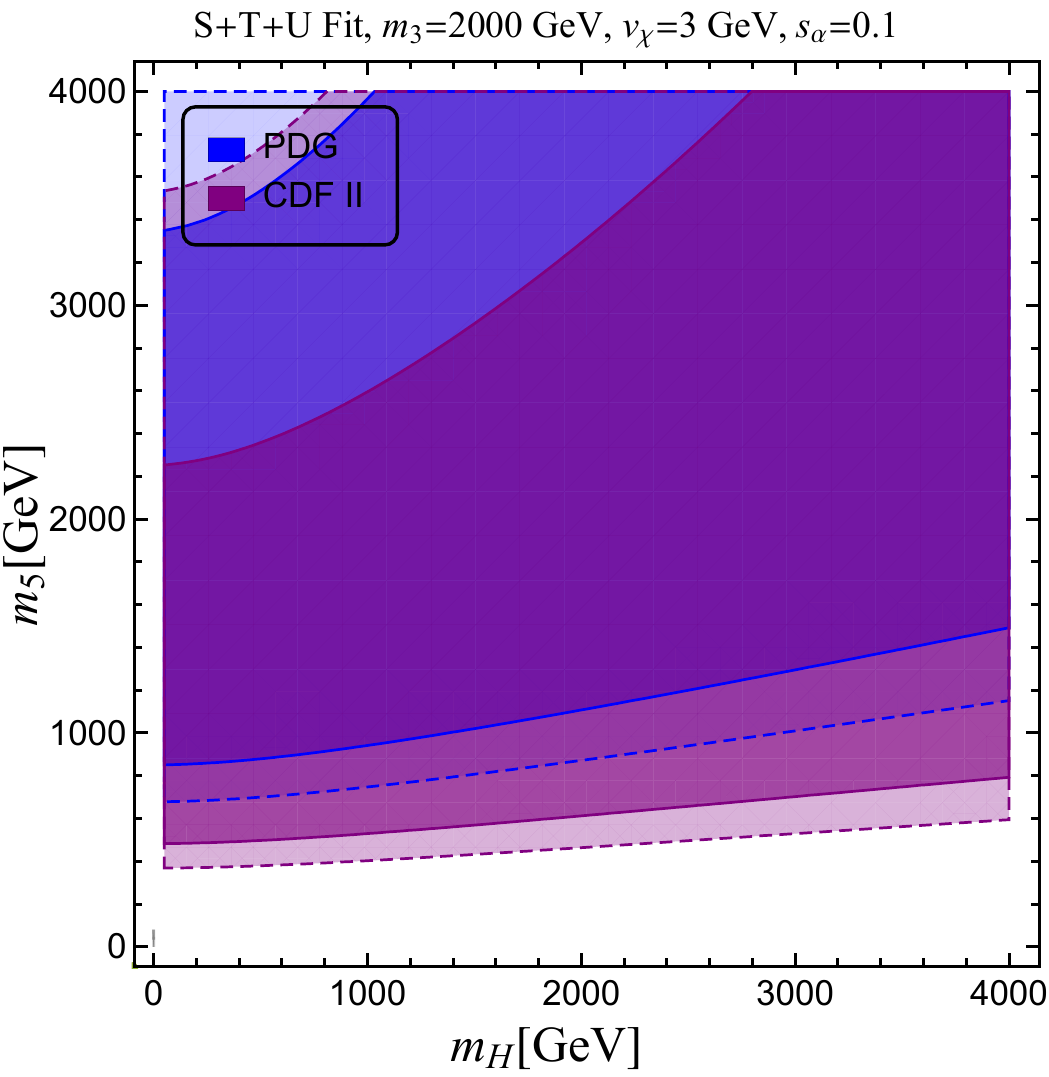}
\end{tabular}
\end{adjustbox}
}
\caption{Same as figure~\ref{plot:realandcomplex} but for the Georgi-Machacek model. See text for details.}\label{plot:GMModel1}
\end{figure}

\begin{figure}[!htp]
\centering{
\begin{adjustbox}{max width = \textwidth}
\begin{tabular}{cc}
\includegraphics[width=0.4\textwidth]{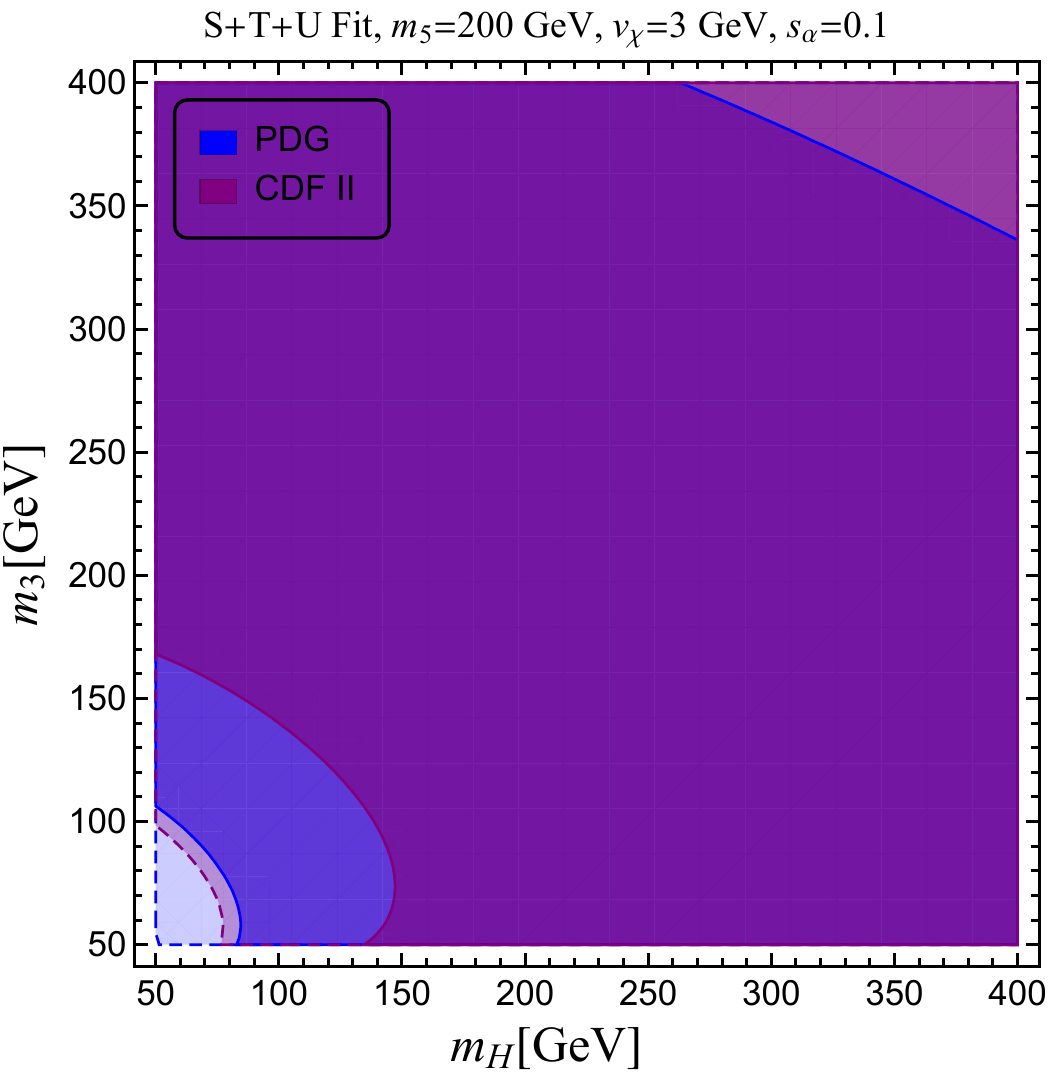} & 
\includegraphics[width=0.4\textwidth]{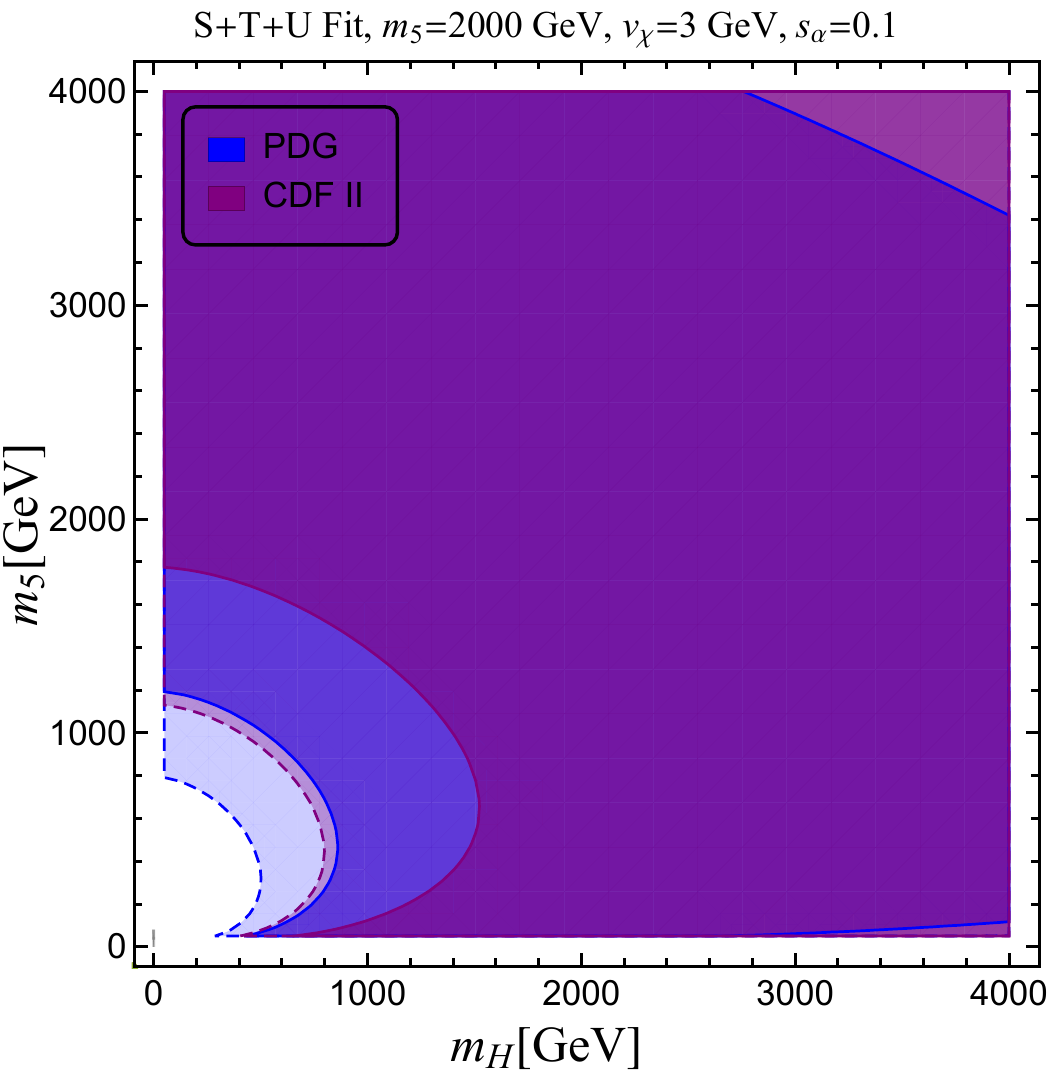} \\ 
\includegraphics[width=0.4\textwidth]{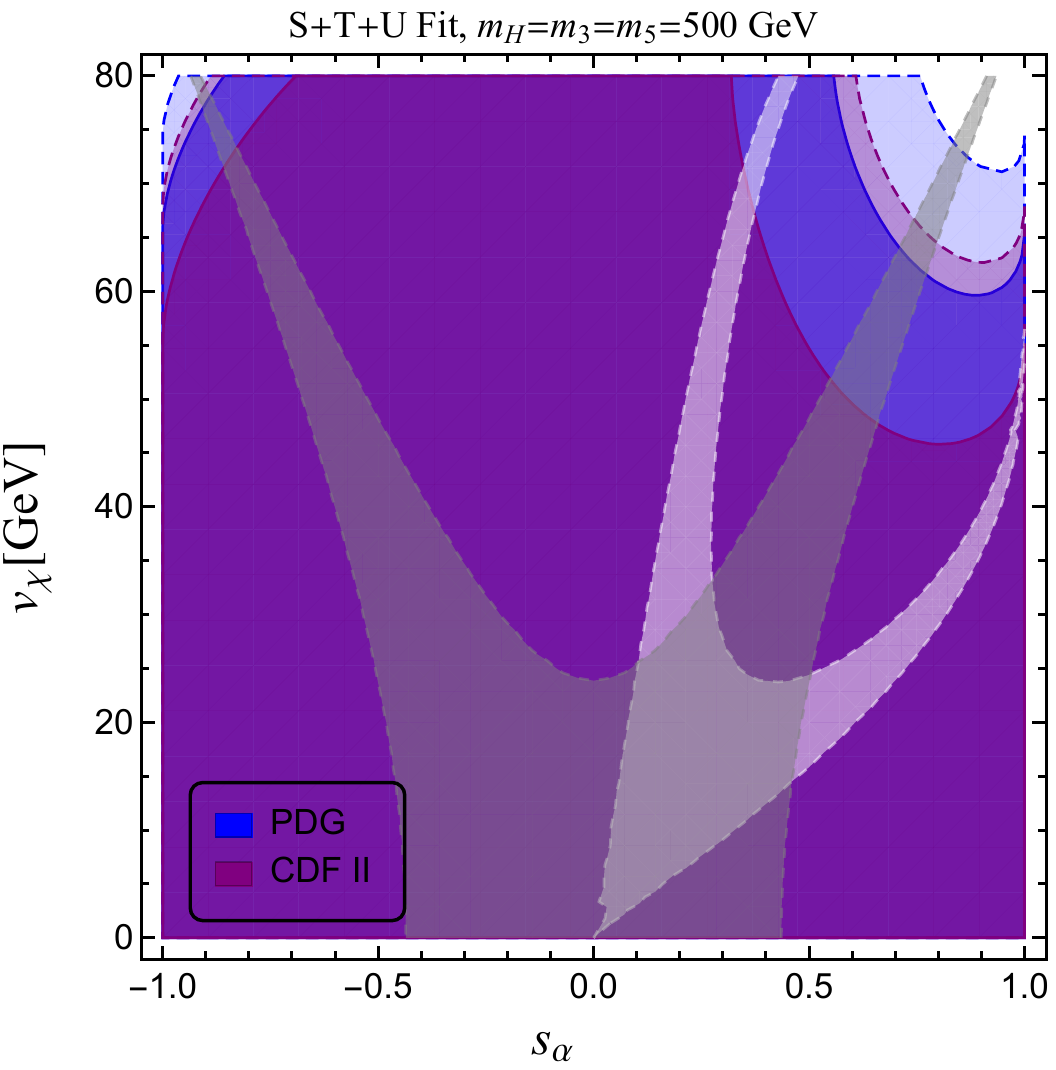} & 
\includegraphics[width=0.4\textwidth]{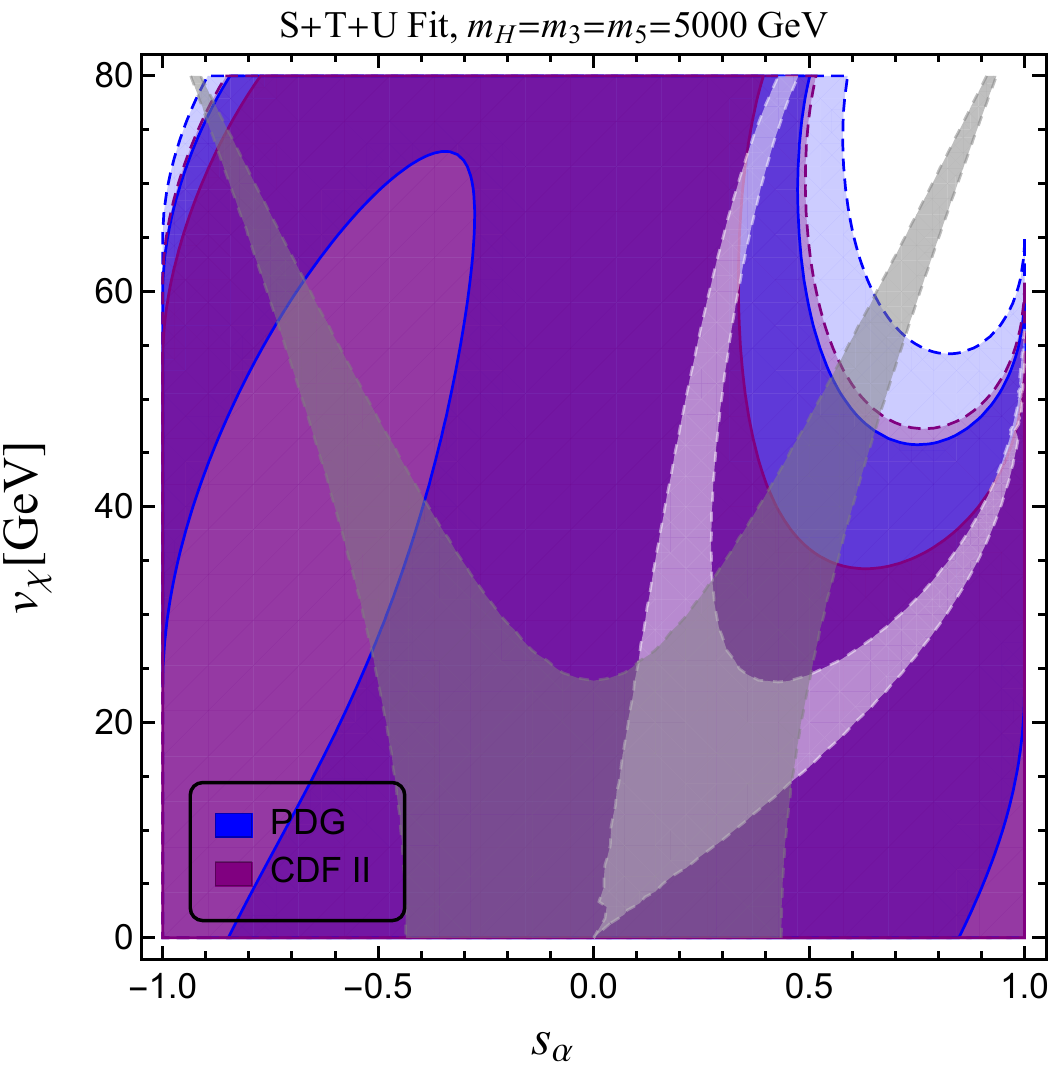}
\end{tabular}
\end{adjustbox}
}
\caption{Figure~\ref{plot:GMModel1} continued. In the lower two panels of $v_\chi-s_\alpha$ plots, we also show the allowed 1-$\sigma$ regions of the ratios of the couplings of the SM-like $h$ to the weak gauge bosons $\kappa_V$ (white) and fermions $\kappa_f$ (Gray) to the corresponding SM values from ATLAS and CMS combined measurements~\cite{Workman:2022ynf}. Such Higgs coupling measurements favor positive values of $s_\alpha$ due to the fact that $\kappa_V=c_\alpha c_\beta+2\sqrt{6}s_\alpha c_\beta/3$ is an asymmetric function of $s_\alpha$~\cite{Kanemura:2013mc, Chiang:2014bia}, therefore we choose $s_\alpha=0.1$ as our benchmark value in other plots. \label{plot:GMModel2}}
\end{figure}

We show the results for the Georgi-Machacek model in Fig.~\ref{plot:GMModel1}-\ref{plot:GMModel2} by marginalization over $T$ and assuming the two triplets have the same VEV to preserve the custodial symmetry $\rm SU(2)_V$ at the tree-level. 
Due to the relatively large number of parameters, we produce plots by projecting onto the $m_5-m_{3,H}$ plane in the plots of Fig.~\ref{plot:GMModel1}, the $m_3-m_H$ and the $v_\chi-s_\alpha$ planes in Fig.~\ref{plot:GMModel2}. All notations in our plots follow those in Ref.~\cite{Gunion:1990dt, Chiang:2012cn, Kanemura:2013mc}. For projections onto the $m_5-m_3$ mass and $m_5-m_H$ mass plane, we find the CDF II data generically prefers a lighter $m_5$ as indicated by the purple regions in Fig.~\ref{plot:GMModel1}, and note that the conclusion remains qualitatively the same when varying the other model parameters. In contrast, as seen from the upper left panel of Fig.~\ref{plot:GMModel2}, with $m_5$ and the other model parameters fixed, the CDF II data would generically prefer a larger $m_{3,\, H}$. In particular, negative mass splittings between $m_{3,\, H}$ and $m_5$ would be relatively disfavored by the CDF II data. 
Relatively large values of $m_{3,\, H}$ can safisfy with the CDF II data, which used to be ruled out by PDG data. Lastly, we show in the rest plots of Fig.~\ref{plot:GMModel2} the sensitivity to the triplet VEV $v_\chi$ and the mixing angle $s_\alpha$ in the degenerated mass spectrum scenario. We find that: 
(1) when the Georgi-Machacek model is at the weak scale, the allowed parameter space would be insensitive to $s_\alpha$ for small $v_\chi$ ($v_\chi\lesssim 40$~GeV), while the CDF II data will rule out slightly more parameter space where $|s_\alpha|$ is close to one and $v_\chi$ is close to $80$~GeV; (2) when the Georgi-Machacek model is above a few TeV, new parameter space, as indicated by, for example, the purple region with the blue boundary, will open up to account for the CDF II result in the negative $s_\alpha$ region; (3) however, the current Higgs coupling measurements (especially the couplings between Higgs and massive gauge bosons) requires a positive $s_\alpha$ under our notations, indicating that the new negative $s_\alpha$ region allowed by the CDF II result is not consistent with Higgs measurements.

\section{Conclusions \label{sec:con}}
While the SM has been very successful and precisely tested, it has also been known for a while that one needs to go beyond it to address several profound problems in nature. These include the baryon asymmetry problem, dark matter, and the non-vanishing neutrino masses. Scalar extensions of the SM have been actively investigated in literature in the past due to its possible ability to the explain two or even all of these three problems simultaneously. More importantly, not all these new scalars below the TeV scale have been fully ruled out at the LHC, thus it will be possible to further test these scalar models at current and/or future colliders. In light of the recent anomaly on $W$ mass measurement from CDF II, we try to address this issue in the frame work of scalar extensions of the SM, and pin down the regions that have been previously thought been ruled out ahead of the CDF II result. To that end, we consider two scenarios in this work: scalar extensions without custodial symmetry ({\textbf{Scenario I}}) and ones with custodial symmetry such as the GM model and its generalizations ({\textbf{Scenario II}}). 

Specially in {\textbf{Scenario I}}, we analyze two cases: {\textbf{Case A}} focuses on the scenario where only the SM doublet can develop a non-vanishing VEV after electroweak spontaneous symmetry breaking, while {\textbf{Case B}} on the scenario where the new scalars can also develop non-vanishing VEVs. To be concrete, we consider a generic SU(2) $N$-plet in {\textbf{Case A}}, realized by imposing an extra discrete $\mathbb{Z}_2$ symmetry. For {\textbf{Case B}}, we consider specifically the real and the complex triplet model due to the fact that relatively light such particles at the $\mathcal{O}(100\,\rm GeV)$ scale are still permitted\,\cite{Du:2018eaw,Chiang:2020rcv}, providing very interesting targets for future collider searches.

Our results for {\textbf{Case A}} are shown in Fig.\,\ref{plot:N-plet}, and Fig.\,\ref{plot:realandcomplex} for {\textbf{Case B}} in {\textbf{Scenario I}}. In either case, we find new parameter space that are previously disfavored prior to the CDF II result. Specifically, in the $N$-plet case, we find the CDF II data would generically prefer larger mass splittings between the multiplet particles. For the real triplet in {\textbf{Case B}}, the conclusion depends on significantly on whether the real triplet develops a non-vanishing VEV or not after the electroweak symmetry breaking: (1) If the real triplet has a vanishing VEV after the symmetry breaking, then CDF II would prefer a heavier real triplet particles above 40~TeV than the PDG data; (2) In contrast, if the real triplet has a non-vanishing VEV after the symmetry breaking, we find while the PDG roughly favors a degenerate spectrum for the real triplet, the CDF II result would prefer a region where the neutral component of the real triplet is heavier than its charged one. On the other hand, for the complex triplet, we find a relatively light complex triplet below the TeV scale is preferred by the CDF II data, making it an interesting target at colliders.

Inspired by the possibility of the lightness of the masses of the triplets and also the custodial symmetric property of the scalar potential, we further consider the Georgi-Machacek model, a mixture of the SM doublet, the real, and the complex triplet scalars, as a particular example of {\textbf{Scenario II}}. Requiring vacuum alignment between the complex and real triplets, pure loop effects in such model can still alleviate the tension between the CDF II data and EWPOs with electroweak scale new particles. A particular hierarchy of the mass spectrum of the GM model is favored by the CDF II data, $m_{3,\, H}>m_5$, while they do not give more constraints on the extra scalar VEV or singlets mixing angle $\alpha$ taking into account the LHC Higgs data.

With the observed CDF II $W$ mass anomaly, it is the time to reconsider the effects of custodial symmetry in the BSM. Our work takes the first step in this direction by analyzing the scalar extensions of the SM. However, there are still a lot of questions unanswered. Especially given a huge enthusiasm for the SMEFT in the community, whether SMEFT is a suitable description of our nature if custodial symmetry is broken by a tree-level VEV is still questionable~\cite{Skiba:2010xn, Khandker:2012zu}. We will leave these for future studies in a subsequent publication.

\paragraph{Note added.} We note that Ref.~\cite{Bahl:2022gqg} studied the compatibility of spectrum preferred by the CDF II $W$ mass with other EWPOs and Higgs precision data in inert complex triplet extension (with hypercharge $Y=2$) of the SM, and Ref.~\cite{Cheng:2022hbo} studied both complex and real triplet extensions by directly perform the global fit to $19$ EWPOs with CDF II $m_W$ via Bayesian analysis. Both references also discussed some related collider phenomenology. Our $STU$ fit results for complex and real triplet qualitatively agree with theirs, while we focus more on the general scalar extensions and also effects from custodial symmetry.

\acknowledgments
We thank Yong Du for his work at the early stage of this project. H.S. also thanks Wei Su for useful discussions. H.S. is supported by the International Postdoctoral Exchange Fellowship Program. X.W. is supported by the Fundamental Research Funds for the Central Universities of China under Grant No.~GK202003018. J.-H.Y. is supported by the National Science Foundation of China under Grants No. 12022514, No. 11875003 and No. 12047503, and National Key Research and Development Program of China Grant No. 2020YFC2201501, No. 2021YFA0718304, and CAS Project for Young Scientists in Basic Research YSBR-006, the Key Research Program of the CAS Grant No. XDPB15.

\bibliographystyle{JHEP}
\bibliography{ref}

\end{document}